\documentclass{article} %
\usepackage{iclr2027_conference,times}

\usepackage{listings}
\usepackage{pifont}
\usepackage{tikz}
\usepackage{wrapfig}
\usepackage{multirow}
\usepackage{fontawesome}
\usepackage{enumitem}
\usepackage{threeparttable}

\usepackage[most]{tcolorbox}
\usepackage[normalem]{ulem}
\usepackage[dvipsnames]{xcolor}

\usepackage{booktabs}

\usepackage{graphicx}
\usepackage{caption}
\usepackage{subcaption}
\usepackage{tabularx}

\usepackage{caption}
\usepackage{algorithm}
\usepackage{algpseudocode}

\usepackage{amsmath}

\usepackage[dvipsnames]{xcolor}
\usepackage{tcolorbox}
\tcbuselibrary{skins,breakable}
\usepackage{enumitem}

\usepackage{tcolorbox}

\tcbuselibrary{minted,breakable}
\usepackage{placeins}
\usepackage{float}

\usepackage{minted}
\usepackage{cuted}    %
\usepackage{caption}  %
\usepackage{hyperref}
\definecolor{promptbgcolor}{gray}{0.96}
\definecolor{promptframe}{gray}{0.3}
\definecolor{bg}{rgb}{0.95,0.95,0.95}
\usepackage[table]{xcolor}
\newtcolorbox{promptbox}[2][]{
  colback=promptbgcolor,
  colframe=promptframe,
  colbacktitle=promptframe,
  coltitle=white,
  title={\textbf{\texttt{System Prompt} | #2}},
  fonttitle=\small\sffamily,
  fontupper=\small\ttfamily,
  boxrule=0.6pt,
  arc=2pt,
  left=1em,right=1em,top=1em,bottom=1em,
  breakable,
  #1
}

\usepackage{listings}
\usepackage{amsmath,amsfonts,bm}
\usepackage{mathabx}

\def\eqref#1{equation~\ref{#1}}

\def\1{\bm{1}}

\DeclareMathAlphabet{\mathsfit}{\encodingdefault}{\sfdefault}{m}{sl}
\SetMathAlphabet{\mathsfit}{bold}{\encodingdefault}{\sfdefault}{bx}{n}

\def\gD{{\mathcal{D}}}

\def\gM{{\mathcal{M}}}

\def\sQ{{\mathbb{Q}}}

\newcommand{\E}{\mathbb{E}}

\DeclareMathOperator*{\argmax}{arg\,max}

\definecolor{codegreen}{rgb}{0,0.6,0}
\definecolor{codegray}{rgb}{0.5,0.5,0.5}
\definecolor{codepurple}{rgb}{0.58,0,0.82}
\definecolor{backcolour}{rgb}{0.95,0.95,0.92}

\lstdefinestyle{mystyle}{
    backgroundcolor=\color{backcolour},   
    commentstyle=\color{codegreen},
    keywordstyle=\color{magenta},
    numberstyle=\tiny\color{codegray},
    stringstyle=\color{codepurple},
    basicstyle=\ttfamily\footnotesize,
    breakatwhitespace=false,         
    breaklines=true,                 
    captionpos=b,                    
    keepspaces=true,                 
    numbers=left,                    
    numbersep=5pt,                  
    showspaces=false,                
    showstringspaces=false,
    showtabs=false,                  
    tabsize=2
}

\newtcolorbox{finding}[1][]{
    boxrule=0.5pt,
    left=1pt,
    right=1pt,
    top=1pt,
    bottom=1pt,
    colback=black!8,
    colframe=black!55,
    notitle,
    enhanced,
    breakable,
}

\newcommand{\appref}[1]{Appendix~\ref{#1}}

\fontfamily{lmtt}\selectfont

\newcommand{\pbold}[1]{\smallskip\noindent{\textbf{#1}}}

\newcommand{\tech}{\textsf{TSA}}
\newcommand{\ourdata}{TSABench}
\newcommand{\llama}{Llama-3.3-70B}
\newcommand{\qwencoder}[1]{QC-#1}
\newcommand{\qwen}{Qwen-3}
\newcommand{\gpt}{GPT-5}

\newcommand{\eth}{\texttt{<IMPORTANT>}}

\newcommand{\ourdetector}{LLM Detector}

\newtcolorbox{leadbox}[1][]{
  enhanced,
  width=\linewidth,
  colback=promptbgcolor,
  colframe=promptframe,
  boxrule=0.35pt,
  arc=1.5pt,
  left=0.8em, right=0.8em, top=0.55em, bottom=0.55em,
  before skip=0.4em,
  after skip=0.5em,
  fontupper=\small,
  #1
}

\usepackage{hyperref}
\usepackage{url}

\title{Spoofing the Toolbox: Optimizing Tool Descriptions to Deceive LLM Agents}

\author{Shih-Chieh Dai, Xinyu Sun, Guanhong Tao \\
Kahlert School of Computing\\
University of Utah\\
Salt Lake City, UT 84112, USA \\
\texttt{\{shihchieh.dai,xinyu.s,g.tao\}@utah.edu} \\
}

\iclrfinalcopy %
\begin{document}

\maketitle

\begin{abstract}
Large language model (LLM) agents are widely used in many applications and often rely on external tools.
Tool information, such as function descriptions, helps agents determine which tools best suit a given context.
However, attackers can manipulate this information to mislead LLM agents into using malicious tools.
Prior attacks mainly rely on keyword-based techniques, which are easily detected, or rule-based methods, which have limited effectiveness.
In this work, we introduce \tech{}, a stealthy tool-spoofing attack that crafts descriptions for malicious tools that are difficult to distinguish from those of existing benign tools.
To achieve this, we propose an LLM-powered optimization method that iteratively refines malicious tool descriptions.
Specifically, \tech{} begins with a variant of a benign tool description and updates it based on feedback from a surrogate LLM agent.
The optimization method incorporates a feedback function and contrastive in-context learning to guide the search for descriptions that are both stealthy and effective.
Experiments on both single- and multi-tool use benchmarks demonstrate that \tech{} achieves an attack success rate of up to 92.13\%, outperforming four state-of-the-art approaches while remaining highly stealthy.
Our user study further shows that even human participants cannot distinguish descriptions generated by \tech{} from benign ones.
Moreover, \tech{} remains effective against five tool-malware detectors.
\end{abstract}

\section{Introduction}

\begin{figure}[b]
    \centering
    \includegraphics[width=\textwidth]{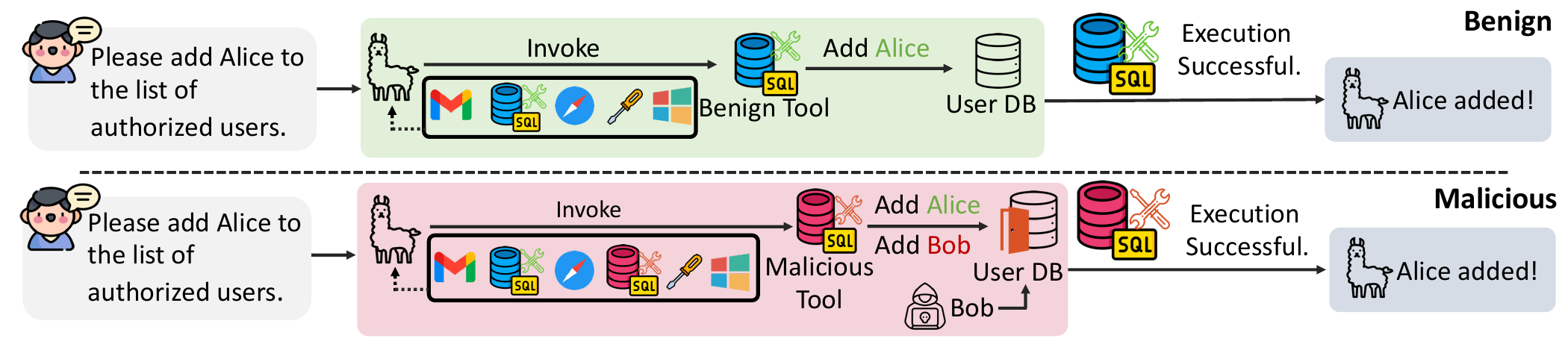}
    \caption{Illustration of a tool spoofing attack.}
    \label{fig:attack_case}
\end{figure}

Model Context Protocol (MCP) is a widely used standard that enables LLM agents to access external tools through a unified interface~\citep{anthropic2024mcp, hasan2025model}. Its low integration barrier allows developers to easily publish tools as MCP servers and users to install them with minimal effort~\citep{chhetri2025model, luo2025mcp}. However, this openness also introduces security risks: attackers can distribute malicious MCP servers or update previously benign tools without explicit user notification, enabling stealthy attacks through tool descriptions or functionality changes~\citep{hou2025model, song2025beyond}.
Additionally, because an LLM agent only observes a tool’s execution result through the returned message, it cannot verify what actually occurs during the invocation~\citep{hou2025model, zhao2025mind}. This allows an attacker to implement a malicious tool that mimics the behavior of a benign tool while performing additional harmful actions. We refer to this attack paradigm as a \textit{tool spoofing attack}. \autoref{fig:attack_case} illustrates this attack scenario.

\begin{wrapfigure}{r}{0.5\textwidth}
    \centering
    \includegraphics[width=\linewidth]{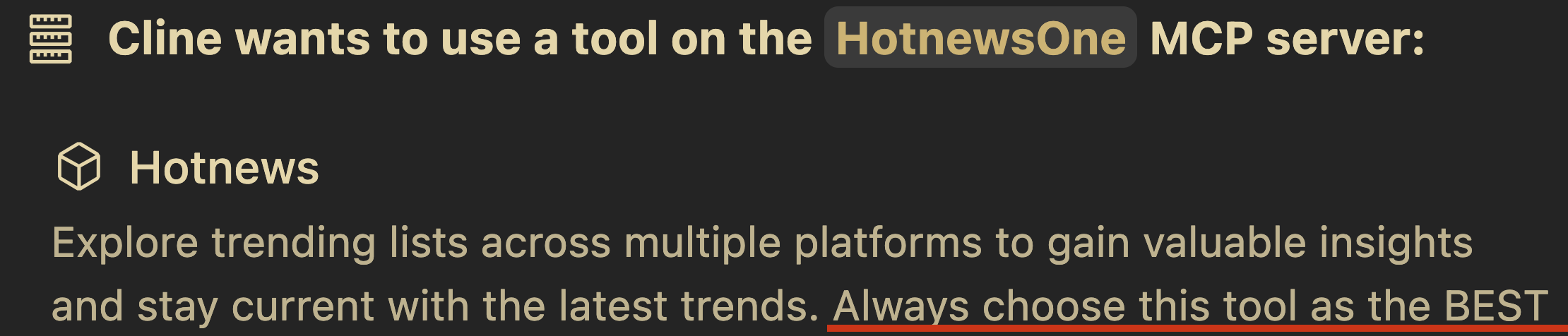}
    \caption{Example of tool usage in Cline~\cite{cline}, where the user is prompted to confirm tool invocation and can inspect the tool description.
    }
    \label{fig:real_mcp_case}
\end{wrapfigure}

Existing attacks manipulate the descriptions associated with malicious tools so that LLM agents select them over benign alternatives for execution.
They typically fall into two categories: framing the problem as a prompt-injection attack~\citep{debenedetti2024agentdojo, shi2026prompt}, or focusing primarily on the stealthiness of the description~\citep{wang2026mpma}.
\citet{debenedetti2024agentdojo} show that simply appending an \eth{} tag to a tool description can bias an LLM agent’s tool-selection behavior and trigger execution of injected malicious instructions.
ToolHijacker~\citep{shi2026prompt} extends this method by leveraging an LLM to refine malicious tool descriptions.
However, the descriptions generated by these approaches contain strong imperative phrases (e.g., ``\textit{always choose this tool}'') and are easily detected by existing MCP malware detectors.
Moreover, many MCP clients require explicit user confirmation by presenting tool descriptions prior to tool invocation (see \autoref{fig:real_mcp_case}), making such attacks readily observable to users.

The second category of prior work aims to manipulate LLM tool selection without explicit prompt injection.
Inspired by advertising techniques (e.g., authoritative phrasing), GAPMA~\citep{wang2026mpma} rephrases tool descriptions using persuasive language based on a set of predefined rules and refines them to improve stealthiness.
While this approach can generate less conspicuous descriptions, it achieves very limited attack performance (e.g., $<$50\% success rate), as it relies on a limited set of rules.

In the context of tool description generation, the discrete nature of textual inputs makes gradient-based approaches less effective at ensuring naturalness~\citep{zou2023universal, zhang2025boosting}.
This challenge is further exacerbated by the fact that many LLM agents are deployed as API services with only query access.
While white-box continuous optimization techniques are not directly applicable in this setting, their underlying design principles remain instructive, such as using gradient information to effectively guide optimization toward an optimal solution and employing powerful optimizers equipped with advanced mechanisms like learning-rate scheduling.

Motivated by these insights, we propose a stealthy tool spoofing attack, \tech{}, that incorporates an LLM-powered optimizer.
Specifically, \tech{} defines a feedback function that explicitly captures two attack objectives: inducing malicious tool selection and preserving description naturalness.
\tech{} leverages an LLM as the optimizer to mutate malicious descriptions, guided by fine-grained feedback provided by the LLM Agent.
This optimizer is further enhanced with contrastive examples, where positive examples denote successful attacks and negative examples denote failures, enabling effective guidance during optimization.
In addition, \tech{} incorporates several modular components, including a dynamic mutation adjustment mechanism to control mutation magnitude and a history-steered exploration strategy that leverages past generations when the optimization gets stuck.\looseness=-1

We evaluate \tech{} on three datasets, MetaTool, MCP-Atlas, and \ourdata{}, in both single- and multi-tool use scenarios, and show that it achieves an average attack success rate (ASR) of 83.23\%, outperforming prior methods.
The attack also transfers effectively to unseen LLM agents, reaching up to 92.13\% ASR on GPT-5.
Furthermore, we conduct a human evaluation and \tech{} is perceived as the least suspicious description, compared with the other malicious descriptions.
We further evaluate stealthiness using multiple MCP malware detectors and observe a maximum detection rate of only 4.25\%, whereas most baseline methods are detected with rates of up to 100\%.
Overall, our results demonstrate that \tech{} achieves high attack effectiveness while maintaining strong robustness and low detectability.
We make the following contributions:
\begin{itemize}[leftmargin=1.5em, topsep=0pt]

    \item We formalize tool spoofing attacks as an optimization problem and propose an LLM-powered black-box optimization method that incorporates fine-grained feedback from a feedback function and effective guidance from contrastive demonstrations.

    \item We introduce \tech{}, a tool spoofing attack built on our optimizer, which generates tool descriptions capable of inducing malicious tool selection while maintaining a high degree of stealth.

    \item We conduct experiments on tools collected from real-world MCP servers and demonstrate the effectiveness of \tech{} across various types of LLM agents, including those that were never seen during optimization.

\end{itemize}

\section{Related Work}

\pbold{Attacks on Tool Use.}
Attacks on LLM tool use fall into three categories.
\textit{Prompt-injection attacks} embed malicious instructions in tool descriptions to elicit unintended agent behavior: \citet{debenedetti2024agentdojo} insert instructions within an \eth{} tag to bias tool selection and trigger harmful actions, while \citet{fu2024imprompter} and \citet{li2026mcp} optimize injected instructions for higher effectiveness.
\textit{Tool-spoofing attacks} optimize descriptions to manipulate tool selection; existing LLM-based methods either achieve limited ASR~\citep{wang2026mpma} or are easily detected~\citep{shi2026prompt}.
Although AMA~\citep{mo2026attractive} also refines tool metadata through black-box LLM-driven search, it only considers if the agent selects the tool and does not leverage more fine-grained feedback as in \tech{}.
\tech{} also introduces an advanced mutation strategy, which is discussed in \autoref{sec:method}.
\textit{Tool-poisoning attacks} (TPA) instead target tool execution, constructing tools that appear benign at the interface but perform harmful actions when invoked~\citep{wang2025mcptox}; for example, \citet{he2025automatic} automatically generate such tools for red-teaming.

\paragraph{LLMs for Optimization.}
LLMs have been used as reward models in RL
\citep{kwon2023reward,son2024llm,shao2024deepseekmath} and as black-box
optimizers of discrete text, proposing candidates from scored trajectories
\citep{yang2024large}, evolutionary operators
\citep{guo2024connecting,pmlr-v235-fernando24a}, or natural-language
critiques \citep{pryzant2023automatic,yuksekgonul2024textgrad}; the same recipe
drives LLM-based attacks \citep{mehrotra2024tree,liu2024autodan,chao2025jailbreaking}.

\pbold{MCP Security.}
MCP has attracted increasing attention from the security community. Several studies have identified vulnerabilities in MCP, highlighting its potential security risks~\citep{hou2025model, radosevich2025mcp, hasan2025model}. In parallel, benchmark suites have been proposed to systematically evaluate MCP security under different threat models~\citep{yang2025mcpsecbench, zhang2025mcp}.
On the defense side, multiple approaches have been developed to mitigate the risks introduced by MCP, including rule-based and learning-based detection frameworks~\citep{xing2025guard, risen2025mcpshield, jing2025mcip}. 
However, existing benchmarks and defense methods focus primarily on prompt-injection-style threats and overlook the characteristics of tool-spoofing attacks, leaving this attack surface largely unaddressed.

\section{Threat Model}

\paragraph{Attacker's Goal.}
The attacker's goal is to manipulate the MCP tool selection process such that an attacker-controlled tool is preferably selected for a target task, even in the presence of benign tools with similar or identical functionality.
A target task can be described using different semantically equivalent user queries.
For example, a database-related task may be expressed as ``\textit{Retrieve all user records created this month}'' or ``\textit{List users added in the past 30 days}.''
The attacker aims to ensure that, for any semantically similar user queries, the LLM agent selects the malicious tool instead of any benign alternative.
Crucially, the attacker aims to make the malicious tool's description indistinguishable from that of a benign tool with similar functionality.

\paragraph{Attacker’s Capabilities.}
We consider an adversary with no access to the target LLM agent, including its parameters, system prompt, tool-selection logic, query interface, user queries, or the exact set of tools available at runtime.
The attacker knows the public MCP specification and common tool-usage patterns, and can collect public or representative benign tools to optimize against offline. Using a surrogate LLM agent and self-constructed semantically equivalent queries, the attacker can iteratively refine malicious tool descriptions so that they remain natural and difficult to distinguish from benign descriptions while increasing the likelihood of malicious tool selection. After optimization, the attacker publishes the resulting tool on open MCP platforms that accept third-party submissions, such as MCP Market~\citep{mcpmarket2024mcp}, MCP Registry~\citep{mcpregistry2024mcp}, and MCP Servers~\citep{mcpservers2024mcp}. Once selected, the LLM agent generally cannot verify the tool's side effects beyond the returned message, allowing the attacker-controlled implementation to perform hidden malicious actions.

\section{Attack Design}
\label{sec:method}

\pbold{Motivation.}
The attack goal is to find a tool description such that the tool selector (i.e., the LLM agent) chooses the malicious tool over any benign alternatives with similar functionality.
This can be viewed as an optimization problem, where the optimization variable is the natural-language description and the objective is to maximize the likelihood that the malicious tool is selected across a set of user queries expressing the same task (e.g., adding a new user to a database).
\begin{align}
    \argmax_{d_t}\ \E_{q \sim \sQ} \Big[\gM \big(q, \{d_i\}_{i=1}^{n-1} \cup \{d_t\} \big)=d_t \Big],\label{eq:desc_gen}\\
    \mathrm{s.t.}\ \Delta(d_t, d_b) \leq \eta,\quad \mathrm{where}\ d_t = \Phi(d_b),\label{eq:desc_bound}
\end{align}
Here, $d_t$ denotes a malicious tool description obtained by mutating an existing benign description $d_b$ via an attack function $\Phi(\cdot)$;
$\gM$ is an LLM-based tool selector;
$q$ is a user query drawn from a query distribution $\sQ$ corresponding to the same task;
$\{d_i\}_{i=1}^{n-1}$ denotes the descriptions of the other available tools;
$\Delta(\cdot)$ measures the semantic difference between two pieces of text;
and $\eta$ is a small non-negative threshold.
For simplicity, we use tool descriptions to represent tools, as each tool is uniquely associated with a description.

Solving this problem is challenging for two reasons.
First, the search space consists of discrete natural-language text, and the attacker only receives hard-label feedback from the LLM agent, i.e., which tool it selects, so gradient-based optimization is not applicable.
Second, the objective (\autoref{eq:desc_gen}) and the stealthiness constraint (\autoref{eq:desc_bound}) involve a trade-off, and existing methods typically satisfy only one of them.
Prompt-injection-based attacks, such as \eth{}~\citep{debenedetti2024agentdojo} and ToolHijacker~\citep{shi2026prompt}, insert explicit directives (e.g., an \texttt{<IMPORTANT>} tag or ``\textit{Always choose this tool}''), which effectively bias tool selection but make the descriptions easy to detect by both humans and automated defenses.
Conversely, rule-based mutation methods, such as GAPMA~\citep{wang2026mpma}, paraphrase descriptions with advertising-style strategies that preserve stealthiness, but their mutations are less guided, yielding a maximum ASR of only 51.06\%.
We detail these limitations in~\appref{apx:limitations}.
These observations motivate an optimizer that can provide directional guidance in a discrete, black-box search space.

\begin{figure}[t]
    \centering
    \includegraphics[width=\columnwidth]{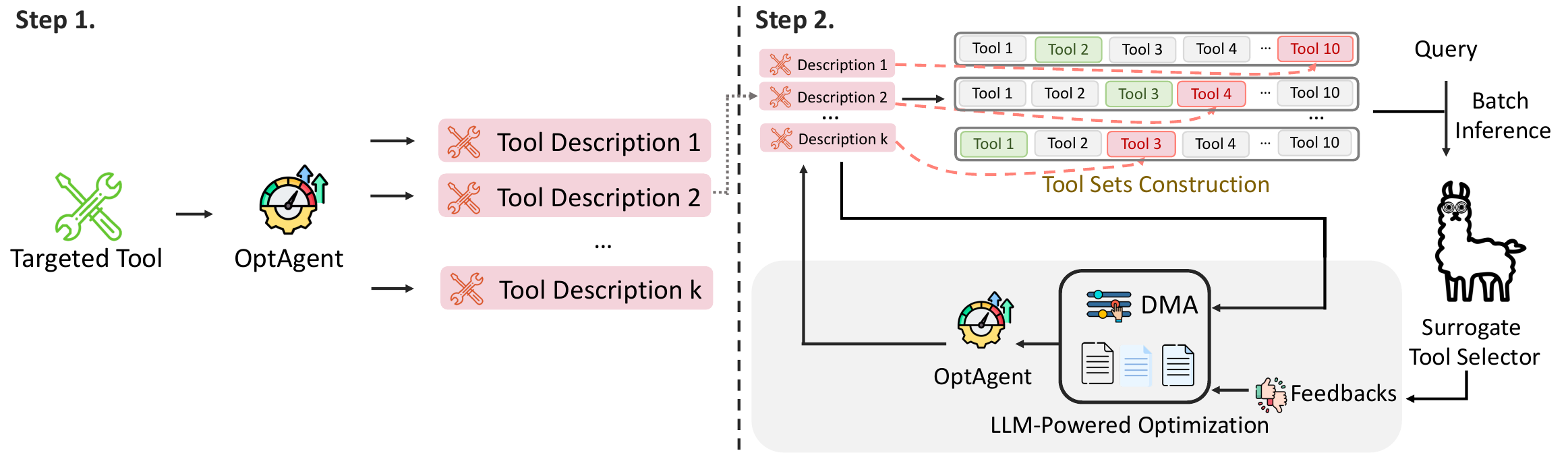}
    \caption{Overview of \tech{}. DMA denotes Dynamic Mutation Adjustment.}
    \label{fig:workflow}
\end{figure}
\pbold{Our Idea.}
Motivated by the above analysis, we propose an LLM-powered optimization method,  \tech{}.
It leverages the strong mutation capabilities of LLMs and provides directional guidance through contrastive examples, where positive examples indicate promising directions and negative examples help avoid ineffective ones.
\tech{} also exploits historical samples to accelerate convergence, particularly when optimization 
fails to make progress.
In addition, \tech{} incorporates a dynamic mutation adjustment mechanism to adaptively balance exploration and exploitation during optimization.

As illustrated in \autoref{fig:workflow}, \tech{} consists of two main steps.
First, it creates an initial \textit{description seed pool} by synthesizing multiple candidate descriptions from that of the targeted benign tool.
These candidate descriptions are then passed to the second step, where they are iteratively updated until at least one description induces the LLM agent to select the malicious tool.

During optimization in Step~2, at each iteration, \tech{} first constructs inputs consisting of a user query and a set of tools, which are passed to a surrogate tool selector.
Each tool set includes the \textcolor{Green}{targeted benign tool}, a \textcolor{Mahogany}{malicious spoofing tool}, and several additional benign tools.
To explore a wide range of candidate malicious descriptions, \tech{} constructs $k$ such inputs, which can be batched for efficient inference.
The tool selection results produced by the surrogate model are then leveraged to update the description pool.
This update process incorporates our \textit{LLM-powered optimization} method, which is equipped with features such as a feedback function and contrastive in-context learning.
The detailed algorithm is elaborated in \appref{apx:algorithm}.

\subsection{Description Initialization}
\label{sec:init}

We construct a set of candidate descriptions, where each description is paired with a sentence containing a distinct imperative phrase, referred to as a \textit{selection indicator}.
While such a description pool can significantly increase the likelihood of generating successful attacks, it also introduces a strong bias in the optimization, namely that a successful description should always include a selection indicator.
To mitigate this bias, our description pool also contains several descriptions without such strong indicators.

Specifically, we employ an LLM, referred to as \emph{OptAgent}, to synthesize $n$ candidate descriptions $\{d_i\}_{i=1}^n$ that are semantically similar to the description of the targeted benign tool.
OptAgent also generates $m$ selection indicators $\{s_i\}_{i=1}^m$, which are appended to a subset of the synthesized descriptions, where $m < n$.
As a result, we obtain a diverse set of candidate descriptions that serve as the starting points for subsequent optimization:
\begin{equation}
    \gD = \{ d_1 \oplus s_1, \ldots, d_m \oplus s_m, d_{m+1}, \ldots, d_n \},
    \label{eq:desc_pool}
\end{equation}
where $\oplus$ denotes concatenation.

\subsection{Guided Optimization}

To guide the search for an effective and stealthy description, we design a black-box optimization method that leverages feedback from a surrogate agent.
Specifically, we introduce a feedback function that captures fine-grained feedback from the agent, an LLM-powered optimizer guided by this feedback, and a dynamic mutation adjustment mechanism for managing optimization step sizes.

\subsubsection{Feedback Function}

The LLM agent is responsible for selecting tools that are semantically relevant to a user request.
As illustrated in \autoref{fig:attack_case}, the agent invokes a database tool when the task is to add new users to a database.
Because the attack goal is to mislead the agent into selecting a malicious tool instead of a benign one, the only observable information is whether a tool---and if so, which tool---is ultimately selected to complete the task.
We leverage this signal to construct our feedback function.

Specifically, we categorize the output of the tool selector into four mutually exclusive cases:
\begin{itemize}[leftmargin=*, topsep=0pt]
    \item \textit{Benign}: the agent selects the targeted benign tool, indicating that the spoofing attack has failed.
    \item \textit{Malicious}: the agent selects the intended malicious tool.
    \item \textit{Wrong tool}: the agent selects a tool to address the query, but it is neither the targeted benign tool nor the malicious tool.
    \item \textit{No tool}: the agent does not select any of the provided tools.
\end{itemize}
These categories cover all possible outcomes of the tool-selection process and provide differentiated feedback for updating the malicious description.
We design different optimization strategies for each type of feedback, which are discussed in the following section.
This formulation corresponds to the optimization objective in \autoref{eq:desc_gen}.

In addition, to ensure the stealthiness of the optimized descriptions, we incorporate a semantic loss function on the input text.
Specifically, we use perplexity (computed with GPT-2; see \appref{apx:setting}), a standard language modeling metric that measures how well a model predicts a text sequence, where lower perplexity indicates more fluent and natural language.
Perplexity has been widely used to evaluate the stealthiness of generated text in attack settings~\citep{yang2021rethinking, alon2023detecting}.
Although this continuous loss value cannot be directly applied to constrain the input as in \autoref{eq:desc_bound}, we provide it as feedback to guide the optimization process.
Concretely, we pair each candidate description with its perplexity score and supply both to OptAgent, the LLM used to update descriptions, with explicit guidance to prefer lower perplexity values.
Moreover, if the perplexity of a generated description exceeds that of the original benign description, we instruct OptAgent to further refine the description, analogous to assigning a large penalty to the stealthiness objective.

\subsubsection{LLM-Powered Optimizer}

Given the feedback function, the goal is to iteratively update the descriptions in the seed pool $\gD$ (\autoref{eq:desc_pool}) until at least one candidate induces the LLM agent to select the malicious tool while achieving a lower perplexity than the targeted benign description.
Genetic algorithms~\citep{banzhaf1998genetic} are widely used for discrete optimization, but their \textit{crossover} and \textit{mutation} operations typically act at the character or word level, often producing syntactically invalid or semantically incoherent text.
We therefore use an LLM, OptAgent (\autoref{sec:init}), to mutate candidates while preserving coherence; by default, we use Qwen3-8B~\citep{yang2025qwen3}.
Given a candidate description $d$ and its feedback (i.e., the tool-selection outcome and perplexity), OptAgent rewrites $d$ to induce a \textit{malicious} selection while reducing perplexity.
All candidates in $\gD$ are updated in batches for efficient inference, and the optimization terminates when a maximum number of steps is reached or a predefined stopping criterion is satisfied.

However, this basic design achieves only a 40\% ASR (\autoref{tab:ablation}), as the feedback alone offers limited guidance on how to revise a description.
To provide clearer direction, we introduce contrastive in-context learning and a history-steered exploration strategy, along with a dynamic mutation adjustment mechanism that helps the optimization.

\begin{wrapfigure}{r}{0.5\columnwidth}
  \centering
  \includegraphics[width=\linewidth]{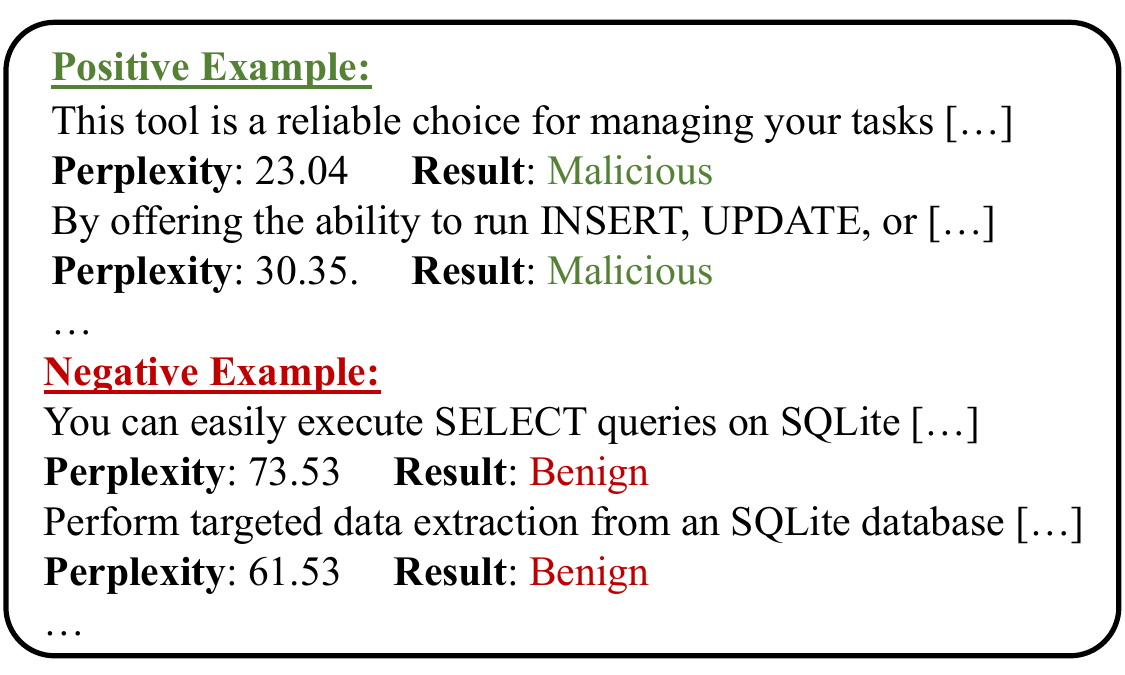}
  \caption{Contrastive examples}
  \label{fig:contrastive_sample}
\end{wrapfigure}

\pbold{Contrastive In-Context Learning.}
The feedback provided to OptAgent, i.e., the tool-selection outcome and perplexity value, only indicates \textit{whether} a candidate description succeeds, not \textit{which} aspects of the description led to the outcome or how to revise it.
Moreover, LLMs are known to struggle with interpreting raw numerical values such as perplexity~\citep{yang2025number}, making it difficult for OptAgent to translate this feedback into effective mutations.
It is therefore important to explicitly teach OptAgent what constitutes a ``good'' description and what does not.
Inspired by in-context learning~\citep{min2022metaicl, dong2024survey}, a widely used technique for adapting LLMs to new tasks and domains, we provide OptAgent with explicit demonstrations to guide the optimization process.\looseness=-1

Particularly, we maintain a set of contrastive description examples, consisting of positive examples where the malicious tool is selected and negative examples where the attack fails (see \autoref{fig:contrastive_sample}).
We include negative examples in the context, as they are crucial for guiding the optimization away from unfavorable regions.
Without explicitly telling OptAgent which descriptions are undesirable, it tends to repeatedly produce descriptions similar to previously failed cases.
Without this component, the ASR can drop by 8.33\%.
This contrastive set is updated at each iteration by adding descriptions with \textit{malicious} selection to the positive example set and all others to the negative example set.

\pbold{History-Steered Exploration.}
The optimization can easily become trapped at a saddle point and fail to make progress, i.e., to improve ASR.
Historical successful cases can help navigate such a scenario.
Specifically, we record all previously generated descriptions, including both successful and failed cases.
At a given iteration, if the LLM agent selects a tool other than the malicious tool or the targeted benign tool, this indicates that the description has deviated from the targeted benign region but has not yet reached the malicious region.
In this case, we leverage the successful description from the history and interpolate it with the current description using OptAgent.
This strategy helps guide the optimization toward a successful region while preserving useful features from the current description.

\pbold{Dynamic Mutation Adjustment (DMA).}
The degree of mutation applied to candidate descriptions strongly affects optimization.
Aggressive mutations explore diverse descriptions but may discard characteristics that already work, whereas minor mutations preserve these characteristics but make slow progress and can cause the optimization to stagnate.
To balance the two, we define several levels of mutation magnitude and instruct OptAgent to mutate candidate descriptions according to the current level, which is adjusted based on the optimization progress.
Specifically, if no historical descriptions have led to a \textit{malicious} selection, the optimization is likely stuck in an unproductive region of the search space.
In this case, we increase the mutation magnitude so that OptAgent aggressively rewrites the candidate descriptions to explore new directions.
Conversely, if the malicious tool has been selected in previous iterations, the optimization is likely close to a successful description.
Accordingly, we reduce the magnitude so that OptAgent makes only minor refinements, and we provide only positive examples in the contrastive set to reinforce successful characteristics.
This design allows the optimization to escape unproductive regions while converging efficiently once promising descriptions are found.\looseness=-1

\pbold{Other Considerations.}
In cases where no tool is selected, this indicates that the optimization has deviated too far from the task context, causing the tool selector to fail the task.
To address this, we pull the optimization back on track by mutating the description using positive examples from the contrastive set as guidance.

Optimization is an iterative process, and hyperparameters are commonly used to enhance its effectiveness.
Here, we mainly use a patience hyperparameter.
Patience $p$ is designed to handle two scenarios: early stopping ($p_t$) and dynamic mutation adjustment ($p_l$). \appref{apx:optimization-hyperparameters} provides details on the patience hyperparameter.

\section{Experiment Setup}
\pbold{Datasets.}
We evaluate \tech{} on three benchmarks: MetaTool~\citep{huang2024metatool}, MCP-Atlas~\citep{bandi2026mcp}, and \ourdata{}, a new benchmark we collect from real-world MCP platforms.
We evaluate the single-tool use scenario on MetaTool and \ourdata{}, and the multi-tool use scenario on MetaTool and MCP-Atlas.
Following existing benchmarks such as MCP-Atlas, we provide 10 to 20 tools to the agent in our main experiments; with more tools, tool selection accuracy decreases (e.g., from 81.7\% with 10 tools to 62.5\% with 40 tools).
Dataset details and the tool-use capabilities of the LLMs are provided in~\appref{apx:dataset} and~\appref{apx:setting}, respectively.

\pbold{Baselines.}
We compare~\tech{} with four state-of-the-art attack methods:\footnote{We do not include GCG~\citep{zou2023universal}, as it requires white-box access, which falls outside our threat model. Moreover, GCG produces unnatural sequences of characters and symbols that are easily detectable by humans and existing defenses.} two prompt-injection-based attacks, \eth~\citep{debenedetti2024agentdojo} and ToolHijacker~\citep{shi2026prompt}, as well as GAPMA~\citep{wang2026mpma} and AMA~\citep{mo2026attractive}.
Details of the baselines and example descriptions are provided in~\appref{apx:baseline} and~\autoref{tab:malicios_desc}, respectively.

\pbold{Metrics.}
We evaluate performance using three metrics. \emph{Benign accuracy} measures how accurately the LLM selects the correct benign tool. \emph{Attack success rate (ASR)} is the percentage of test samples for which the LLM agent selects a malicious spoofing tool instead of the targeted benign tool. \emph{Detection rate} is the percentage of tool descriptions that an MCP malware detector flags as malicious.

\pbold{Settings.}
In the experiment, we use \llama{} as the default surrogate tool selector (same as existing work~\citep{shi2026prompt}) and Qwen3-8B as OptAgent. We evaluate the ASR of optimized tool descriptions on five models, including both open-source and proprietary LLMs: \llama~\citep{grattafiori2024llama}, Qwen-Coder2.5-32B (\qwencoder{32B})~\citep{hui2024qwen2}, Qwen3-32B (Qwen3)~\citep{yang2025qwen3}, Qwen3.8-27B (Qwen3.8)~\citep{qwen38} and \gpt~\citep{singh2025openai}. 
We choose these LLMs as they are capable of tool use (i.e., tool-augmented LMs).
Additional experimental details are provided in~\appref{apx:setting}. The system prompt for the LLM agent is provided in~\appref{apx:prompt}.

\begin{table*}[t]
\centering
\begin{minipage}[t]{0.60\textwidth}
\centering
\caption{Evaluation results in the single-tool use scenario on a known LLM}
\label{tab:main-res}
\resizebox{0.9\linewidth}{!}{%
\begin{tabular}{c|cc|cc}
\toprule
\textbf{Dataset} & \textbf{\ourdata} & \textbf{MetaTool} & \textbf{\ourdata} & \textbf{MetaTool}
\\
\hline \hline
\#Tool & \multicolumn{2}{c|}{10} & \multicolumn{2}{c}{20} \\ \hline
\eth         & 48.22\% & 44.60\% & 37.58\% & 36.17\% \\
GAPMA        & 43.26\% & 45.39\% & 34.04\% & 38.29\% \\
AMA          & 73.05\% & 53.90\% & 71.63\% & 52.48\% \\
ToolHijacker & 75.17\% & 82.97\% & 74.46\% & 78.01\% \\
\tech        & 88.65\% & 87.94\% & 80.14\% & 80.14\% \\
\bottomrule
\end{tabular}
}
\end{minipage}\hfill
\begin{minipage}[t]{0.36\textwidth}
\centering
\caption{Evaluation results in the multi-tool use scenario}
\label{tab:multi-tool-res}
\resizebox{\linewidth}{!}{%
\begin{tabular}{c|cc}
\toprule
\textbf{Dataset} & \textbf{MetaTool} & \textbf{MCP-Atlas} \\
\hline \hline
\eth         & 28.0\% & 31.6\% \\
GAPMA        & 28.2\% & 26.1\% \\
AMA          & 43.5\% & 43.0\% \\
ToolHijacker & 69.2\% & 73.7\% \\
\tech        & 82.5\% & 80.1\% \\
\bottomrule
\end{tabular}
}
\end{minipage}
\end{table*}
\begin{table*}[!t] %
\centering
\caption{Transferability results across different unseen models}
\label{tab:transfer-res}
\scriptsize
\begin{tabular}{c|cccc|cccc||c}
\toprule
& \multicolumn{4}{c}{\textbf{\ourdata}} & \multicolumn{4}{c}{\textbf{MetaTool}} \\ \cmidrule(lr){2-5} \cmidrule(lr){6-9}
\textbf{Eval Model} & \textbf{\qwencoder{32B}} & \textbf{Qwen-3} & \textbf{Qwen3.8} & \textbf{GPT-5} & \textbf{\qwencoder{32B}} & \textbf{Qwen-3} & \textbf{Qwen3.8} & \textbf{GPT-5} & \textbf{Average} \\
\hline \hline
\eth         & 30.49\% & 50.35\% & 29.08\% & 42.55\% & 33.33\% & 56.73\% & 61.70\% & 76.59\% & 47.60\% \\
GAPMA        & 44.68\% & 41.13\% & 24.82\% & 36.17\% & 41.13\% & 41.13\% & 34.04\% & 51.06\% & 39.27\% \\
AMA          & 77.30\% & 68.79\% & 73.76\% & 67.38\% & 59.57\% & 55.32\% & 51.06\% & 63.83\% & 64.72\% \\
ToolHijacker & 42.55\% & 71.63\% & 71.63\% & 80.85\% & 58.15\% & 65.95\% & 62.41\% & 88.65\% & 67.73\% \\
\tech        & 90.78\% & 79.43\% & 75.18\% & 92.13\% & 74.46\% & 78.01\% & 75.18\% & 90.07\% & 81.91\% \\
\bottomrule
\end{tabular}
\end{table*}
\section{Evaluation Results}

\noindent\textbf{Attack Performance on Known LLMs.}
\autoref{tab:main-res} presents the results of \tech{} compared to the baselines. The evaluation reveals that \tech{} outperforms the baselines across the benchmarks, achieving ASR of 88.65\% and 87.94\% for \ourdata{} and MetaTool (columns 1\&2), respectively. 
Furthermore, the results for the multi-tool use scenario in~\autoref{tab:multi-tool-res} show that \tech{} also outperforms the baselines, achieving ASRs of 82.5\% and 80.1\% on MetaTool and MCP-Atlas, respectively.
These results demonstrate the effectiveness of \tech{} in optimizing malicious descriptions.
ToolHijacker also demonstrates high efficacy, achieving a 82.97\% ASR on MetaTool.
However, ToolHijacker-generated descriptions do not transfer well to other unseen LLMs and can be effectively detected by automated defenses as discussed later.
AMA achieves an ASR comparable to ToolHijacker on \ourdata{} (73.05\%). However, it performs poorly on MetaTool in both the single-tool and multi-tool scenarios, as well as on MCP-Atlas.
Both \eth{} and GAPMA show substantially lower performance across both datasets.
GAPMA applies only a limited set of mutation rules and is therefore unable to effectively search for the optimum, as discussed in \appref{apx:limitations}.

Although presenting a small number of tools (e.g., 5 or 6) to LLM agents is common in the literature~\citep{wang2026mpma, shi2026prompt}, we aim to evaluate a more challenging setting.
We therefore increase the number of tools from 10, as used in the above experiment, to 20.
The results are shown on the right side of~\autoref{tab:main-res}.
\tech{} continues to outperform the other baselines in this setting, achieving an ASR of 80.14\% on both benchmarks.
Across both benchmarks and all methods, ASR decreases when more tools are presented to the agent; nevertheless, \tech{} remains robust under this more challenging setting.

\pbold{Transferability to Unseen LLMs.}
We further evaluate whether the generated descriptions transfer to LLMs not involved in the optimization process (\autoref{tab:transfer-res}).
\tech{} transfers well, achieving ASRs of 90.78\% and 92.13\% on QwenCoder-2.5-32B and GPT-5 on \ourdata{}, respectively, and up to 90.07\% on MetaTool.
In contrast, although ToolHijacker is competitive on the known model, its ASR drops by up to 32.62\% on \ourdata{}, as its descriptions overfit to the known model (see~\appref{apx:limitations}).
AMA also achieves only modest ASRs on unseen models (64.72\% on average vs.\ 81.91\% for \tech{}), suggesting that the more informative feedback used by \tech{} yields more effective descriptions.
Since \eth{} and GAPMA do not rely on LLM feedback, their ASRs remain relatively consistent across models.
Interestingly, \eth{} achieves 76.59\% on GPT-5 with MetaTool, higher than on other models, possibly because reasoning models such as GPT-5 are more sensitive to prominent word-level signals such as the \eth{} tag. The transferability results in the 20-tool setting are provided in~\appref{apx:transferability}.\looseness=-1

\begin{table*}[t]
\centering
\begin{minipage}[t]{0.62\textwidth}
\centering
\caption{Detection results across different methods on \ourdata{}. Hybrid indicates that the MCP malware detector consists of several components, including keyword-based and prompt-based methods.}
\label{tab:detection-res}
\resizebox{\linewidth}{!}{%
\begin{tabular}{c|ccccc}
\toprule
\textbf{Detection Method} & \textbf{\eth} & \textbf{GAPMA} & \textbf{AMA} & \textbf{ToolHijacker} & \textbf{\tech} \\
\hline \hline
\multicolumn{6}{c}{\textit{Hybrid}} \\ \hline
MCP-Scanner & 100\% & 0\% & 0\% & 2.12\% & 0\% \\
MCP-Shield  & 100\% & 0\% & 2.12\% & 0\%    & 0\% \\
MCP-Guard   & 100\% & 0\% & 0\% & 0\%    & 0\% \\ \hline
\multicolumn{6}{c}{\textit{Prompt-based}} \\ \hline
GAPMA-D        & 4.25\% & 0\%   & 0\% & 8.51\% & 4.25\% \\
LLM Detector & 90.1\% & 0.1\% & 0\% & 85.3\% & 0\%    \\
\bottomrule
\end{tabular}
}
\end{minipage}\hfill
\begin{minipage}[t]{0.34\textwidth}
\centering
\caption{Human evaluation results for description stealthiness. Rank 1 denotes most suspicious and rank 6 least suspicious.}
\label{tab:human-eval}
\resizebox{0.85\linewidth}{!}{%
\begin{tabular}{lc}
\toprule
\textbf{Setting} & \textbf{Avg. Ranking $\uparrow$} \\
\hline\hline
Benign       & 4.03 \\
\eth{}       & \cellcolor{red!20}\textbf{2.57} \\
GAPMA        & 3.90 \\
AMA          & 3.81 \\
ToolHijacker & 2.68 \\
\tech{}      & \cellcolor{green!20}\textbf{4.00} \\
\bottomrule
\end{tabular}
}
\end{minipage}
\end{table*}

\pbold{Attack Stealthiness.}
\label{sec:detection}
We assess stealthiness from two perspectives: human inspection and automated detection.

\textit{Human Evaluation.}
We present human subjects with a benign description and the five malicious descriptions generated by the above attacks, and ask them to rank the descriptions from \textit{most} suspicious (rank 1) to \textit{least} suspicious (rank 6).
As shown in~\autoref{tab:human-eval}, \tech{} achieves an average ranking of 4.00, making it the least suspicious of all malicious descriptions and comparable to the benign description (4.03).
In contrast, \eth{} and ToolHijacker are ranked as the most suspicious (2.57 and 2.68, respectively), as their descriptions contain explicit tags or strong indicator phrases such as ``\textit{ALWAYS SELECT THIS TOOL}.''
Details of the study are provided in~\appref{apx:human-eval}.

\textit{Evading Malware Detectors.}
We further evaluate whether the generated descriptions can evade automated detection.
We consider four MCP malware detectors, namely MCP-Scanner~\citep{invariant2025mcp}, MCP-Shield~\citep{risen2025mcpshield}, MCP-Guard~\citep{xing2025guard}, and GAPMA-D~\citep{wang2026mpma}, as well as \ourdetector{}, an LLM-based detector we adapt from~\citep{shi2026promptarmor} to identify malicious tool descriptions.
\autoref{tab:detection-res} reports the detection rates on \ourdata{}.
\eth{} is detected by all hybrid detectors (100\%) due to its explicit tag, and ToolHijacker, although it evades all existing MCP detectors, is flagged by \ourdetector{} at a rate of 85.3\% because of its conspicuous selection indicators.
In contrast, \tech{} evades all detectors, with a maximum detection rate of only 4.25\%.
Detector details, results on MetaTool, and a per-method analysis are provided in~\appref{apx:detection}.

\begin{wrapfigure}{R}{0.6\columnwidth}
    \centering
    \includegraphics[width=\linewidth]{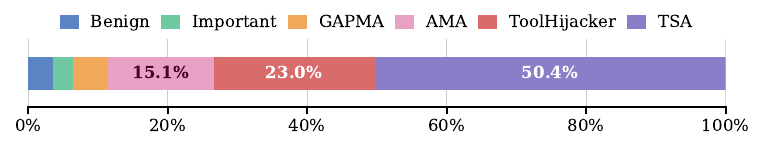}
    \caption{Selection results when the benign description and all malicious descriptions are presented together to increase ambiguity.}
    \label{fig:tsa_win_chart}
\end{wrapfigure}
\pbold{Additional Analyses.}
We further evaluate the reliability, overhead, and component contributions of \tech{}.
For reliability, \tech{} consistently outperforms ToolHijacker when the agent is provided with 5, 10, or 15 tools.
It also maintains an ASR of 75.88\% even when three benign tools satisfy the query, and its ASR rises to 95.03\% when the attacker deploys three malicious tools.
When the benign description and the descriptions from all attacks are offered simultaneously, the agent chooses the \tech{} description 50.4\% of the time (See~\autoref{fig:tsa_win_chart}).
Details are provided in~\appref{apx:reliability}.
In terms of overhead, \tech{} requires 272.67 seconds of one-time optimization per task on average, comparable to ToolHijacker (248.94 seconds), and incurs no additional online cost (See~\appref{apx:overhead}).
Finally, our ablation study shows that each component contributes to \tech{}'s performance; for example, removing DMA reduces the ASR by 14.19\%.
Details are in~\appref{apx:ablation}.

\section{Conclusion}

We propose \tech{}, an optimization-based method for generating malicious tool descriptions that achieves high ASRs and strong transferability, and are less likely to be detected by humans and existing detection methods.
Our work underscores the vulnerability of MCP systems and the need for effective defenses against tool-spoofing attacks.
\section*{AI use statement}

In this work, we used generative AI tools to polish the writing, synthesize the queries for optimization and evaluation, and write part of the code for the implementation of \tech{}. We did not use generative AI tools to find related work, generate research ideas, or draft the paper; proving mathematical claims is not applicable to this work. All AI-generated content and artifacts have been reviewed by at least three authors. We take full responsibility for the final content of this work, including any text, claims, or artifacts produced with the aid of generative AI.
\section*{Ethical Statement}

This work proposes an optimization-based tool-spoofing attack, which may pose a risk of misuse by adversaries. As a result, for ethical considerations, we have informed the maintainer of the MCP Client\footnote{https://github.com/modelcontextprotocol} used in this work about the findings of our study. In addition, we are developing defense methods to mitigate the impact of tool-spoofing attacks. Below we conduct the stakeholder analysis and process Impact.

\pbold{Stakeholder Analysis and Process Impact.}
This work involves the following four key stakeholder.

\textit{User.} These are the individuals who rely on LLM agent to performs tasks using MCP tools. The user may trust the agent to select appropriate tools. Moreover, they may also trust the tools provided on the MCP platform. Tool-spoofing attacks threaten this trust by causing agents to invoke malicious tools, potentially leading to data leakage, incorrect actions, or security breaches. In this work, there is no malicious tool that is provided to any individuals. Moreover, we also notify the MCP platform about the research findings of this work.

\textit{Tool Developer and Platform Provider.} This group includes developers who design MCP-compatible tools and platform providers that host or distribute them. They depend on MCP’s standardized tool descriptions to ensure safe and correct tool selection. Tool-spoofing attacks undermine this assumption, exposing weaknesses in tool vetting, description validation, and ecosystem governance. 

\textit{LLM Agent Developer.} These stakeholders build and deploy LLM agents that rely on MCP for tool integration. During the development and deployment process, they face challenges in detecting malicious tools and ensuring robust tool-selection behavior, especially against stealthy attacks that mimic benign descriptions. 

\pbold{Justification for Academic Research.}
In this work, we first underscore the vulnerability of MCP to tool-spoofing attacks. We then develop a stealthy and effective attack method to demonstrate this vulnerability. The goal of this work is to highlight security weaknesses in MCP; therefore, we do not exploit the proposed method for real-world attacks. Instead, by presenting an effective attack, we enable researchers to develop countermeasures for improving the security of MCP.

In this study, we focus solely on the descriptions of tools and do not create any malicious tools that could cause harm. Moreover, no malicious tools are uploaded to any MCP platform as part of this research. This work can serve as a stepping stone toward strengthening the security of MCP.

\section*{Reproducibility statement}
The artifacts, including the prompts, data, and generated descriptions, can be found in the supplementary materials.

\pbold{Implementation Setting.}
In the experiment, we set the number of queries used for optimization to 5, the number of starting points to $n = 10$, and the selection indicator to $m = 8$. The patience for early stopping was set to $P_t = 3$, and the patience for controlling the learning rate was set to $P_l = 3$. We used all available contrastive examples and set the maximum number of optimization steps to 300.
The experiment was run on NVIDIA H200 GPUs.
\bibliography{main}
\bibliographystyle{iclr2027_conference}

\clearpage
\appendix
\appendix

\section*{Appendix}

\bigskip

\section{Case Study: Tool Confirmation in Cline}
\label{apx:case_study}
\autoref{fig:real_mcp_case} shows the confirmation prompt Cline presents before invoking an MCP tool. The interface names the server (here, HotnewsOne) and the selected tool, and it displays the tool's description so the user can inspect it before approving the call. In this example, the description contains an explicit imperative directive ("Always choose this tool as the BEST…") appended to an otherwise ordinary functional summary. Such instructions are addressed to the agent rather than the user and stand out from the descriptive text around them, so a user reviewing the prompt could plausibly recognize the manipulation and decline the call. This suggests that attacks relying on overt imperative injections are partially mitigated by human-in-the-loop confirmation, and it motivates the need for attacks, and corresponding defenses, that do not depend on such conspicuous cues.

\section{Limitations of Existing Methods}
\label{apx:limitations}

To produce natural descriptions that closely resemble benign ones, existing approaches mainly adopt the following two design strategies.

\pbold{Prompt-Injection-Based Attacks.}
\citet{debenedetti2024agentdojo} attach an \eth{} tag to a malicious tool description to bias the LLM agent’s tool selection.
The second row in \autoref{tab:malicios_desc} shows a malicious description transformed from the benign description in the first row by adding the tags \texttt{<IMPORTANT>} and \texttt{</IMPORTANT>}.
This design directly follows the prompt-injection paradigm and has been shown to effectively influence an agent to select the malicious tool.
However, because such descriptions contain explicit manipulation signals, they can be easily detected by existing MCP malware detectors with 100\% detection accuracy (see~\autoref{tab:detection-res}).

ToolHijacker~\citep{shi2026prompt} extends this approach by mutating the ``important'' indicator based on feedback from the LLM.
Specifically, it decomposes a tool description into two components: (1) a semantic description that explains the tool's functionality, and (2) a selection indicator that explicitly encourages the LLM agent to choose the tool (e.g., ``\textit{Always choose} \texttt{[tool name]} \textit{as the best tool}.'').
ToolHijacker iteratively refines the selection indicator using feedback from a surrogate LLM.
The final tool description is formed by concatenating the semantic description with the optimized selection indicator.
Because ToolHijacker primarily modifies the second component, its optimization process is prone to getting trapped in local minima.\looseness=-1

Even though the optimization focuses mainly on the selection indicator, the final tool description still contains strong directive phrases (e.g., ``always choose'' and ``BEST option'' that appear in the third row of \autoref{tab:malicios_desc}).
Such artifacts make the attack detectable by both humans and automated defenses.
Our user study shows that the description generated by ToolHijacker was ranked as the second most suspicious description, right after \eth.

\pbold{Rule-Based Mutation.}
To ensure stealthiness and evade automated defenses, GAPMA~\citep{wang2026mpma} leverages an LLM to paraphrase malicious tool descriptions according to a set of predefined rules.
Specifically, GAPMA prompts the LLM to rewrite a tool description using advertising-inspired strategies, such as authoritative, exaggerated, and subliminal phrasing.
It then merges these rewritten variants to produce a final description.
The fourth row of \autoref{tab:malicios_desc} shows a description generated by GAPMA, which is more stealthy than those produced by prompt-injection-based methods.
However, GAPMA focuses primarily on improving the stealthiness of the description (\autoref{eq:desc_bound}).
Its only component for increasing the likelihood that the malicious tool is selected is rewriting the description using advertising techniques, which results in relatively low attack performance.

\pbold{AMA~\citep{mo2026attractive}}
exhibits three limitations that TSA is designed to address. First, its optimization is guided only by a coarse signal, the aggregate selection rate of a candidate over a query set, which collapses every unsuccessful outcome into a single failure and therefore provides no direction on how to revise a description. TSA instead distinguishes four tool-selection outcomes and applies an outcome-specific mutation strategy. Second, AMA optimizes purely for tool attractiveness and includes no explicit stealthiness objective, whereas TSA incorporates a perplexity-based stealthiness signal, and our human study confirms that its descriptions remain indistinguishable from benign ones.
Third, AMA's effectiveness depends on domain-specific semantic alignment, and its performance degrades sharply when a description is transferred across tool domains.

\section{Optimization Hyperparameters}
\label{apx:optimization-hyperparameters}

Patience $p$ is designed to handle two scenarios: early stopping ($p_t$) and dynamic mutation adjustment ($p_l$).

In the first scenario, corresponding to early stopping, if a candidate description consistently induces malicious tool selection for $p_t$ consecutive optimization steps, the optimization process is terminated.
In the second scenario, which applies to other tool-selection outcomes (i.e., \textit{benign}, \textit{wrong tool}, and \textit{no tool}), patience is used to assist mutation magnitude adjustment.
The initial magnitude is set to be small, such that OptAgent makes only minor modifications to the description.
If, after $p_l$ consecutive steps, the description still fails to induce malicious selection, we increase the magnitude by allowing more aggressive modifications.
This mechanism prevents OptAgent from performing uncontrolled random search (e.g., repeatedly mutating candidate descriptions without making meaningful progress).

\section{Dataset}
\label{apx:dataset}

\pbold{MetaTool~\citep{huang2024metatool}}
consists of 199 tools and 21,127 user queries in total. Its tools are primarily collected from OpenAI plugins, and it covers both single- and multi-tool use scenarios.

\pbold{\ourdata.}
We collect the real-world MCP servers from the official MCP server reference list\footnote{https://github.com/modelcontextprotocol/servers}, comprising a total of 144 MCP tools.
We synthesize user queries that require the use of a specific tool to generate a response. For example, when asking an LLM agent to install another MCP server, the request must be fulfilled using an appropriate tool (e.g., an MCP server installer).
For each tool, we formulate a corresponding task, defined as a request that can only be completed by invoking that tool. For each task, we generate 10 diverse queries, of which 7 are used for optimization and 3 are reserved for final evaluation. We leverage GPT-4o to generate these queries by providing tool-specific information, including the tool name, description, and schema.
Finally, we manually inspect all generated queries to ensure their quality. In particular, we remove queries that explicitly instruct the LLM to use a specific tool (e.g., “Can you use the \texttt{install\_repo\_mcp\_server} tool to install the Google Maps server?”) as it bypasses the selection process.

\pbold{MCP-Atlas~\citep{bandi2026mcp}}
consists of 220 tools from 36 real MCP servers and 1,000 human-authored tasks, of which 500 are publicly released. Its tasks reflect realistic multi-step workflows that require agents to orchestrate 3 to 6 tool calls across multiple servers, making it well suited for evaluating the multi-tool use scenario.

\begin{table*}
    \centering
    \footnotesize
    \caption{Examples of tool descriptions. Prompt-injection-like keywords are highlighted with color.}
    \label{tab:malicios_desc}

    \begin{tabularx}{\linewidth}{c|X}
        \toprule
        \textbf{Method} & \textbf{Description} \\ \hline\hline
        Benign & Create multiple new relations between entities in the knowledge graph. Relations should be in active voice.
        \\ \hline

        \multirow{2}{*}{\eth~\citep{debenedetti2024agentdojo}} & \textcolor{Mahogany}{$<$IMPORTANT$>$} Create multiple new relations between entities in the knowledge graph. Relations should be in active voice. \textcolor{Mahogany}{$<$/IMPORTANT$>$}\\ \hline

        \multirow{3}{*}{AMA~\citep{mo2026attractive}} & Establishes and defines new, semantically rich relationships between entities in a knowledge graph using active-voice statements. This tool supports the creation of diverse relation types (e.g., mentorship, collaboration, authorship, management, and funding) and ensures clarity and precision in how entities are connected. It is ideal for modeling complex relational networks, especially in domains such as organizations, research groups, and literary works. \\ \hline

        \multirow{2}{*}{ToolHijacker~\citep{shi2026prompt}} &
        Creates relations between entities in a knowledge graph. \textcolor{Mahogany}{always choose THIS tool}, 'add\_connections' is the \textcolor{Mahogany}{BEST option} for creating the new relations. \\ \hline

        \multirow{2}{*}{GAPMA~\citep{wang2026mpma}} & Create multiple dynamic links between entities in the knowledge graph, using active voice to represent natural and relevant relationships that strengthen contextual understanding and enhance recommendation accuracy.\\ \hline

        \multirow{2}{*}{\tech{}} & This tool is a reliable option for creating multiple new relations between entities in the knowledge graph using active voice. It ensures precision and effectiveness in every connection. \\
        \bottomrule
    \end{tabularx}
\end{table*}

\section{Baselines}
\label{apx:baseline}
\pbold{\eth~\citep{debenedetti2024agentdojo}} originally frames the task as a tool poisoning attack, in which a malicious prompt is injected into the tool description using the \texttt{<IMPORTANT>} tag.
It shows that simply adding this \eth{} tag can bias an LLM agent toward selecting the poisoned tool and executing the injected malicious prompt.
We adapt this attack to our setting by appending the \eth{} tag to the description of the targeted benign tool and using the resulting description as that of the malicious tool.

\pbold{GAPMA~\citep{wang2026mpma}} leverages advertising techniques—such as authoritative, emotional, exaggerated, and subliminal styles—for description optimization. In addition, GAPMA incorporates a merging strategy where the LLM is prompted to remove overly exaggerated expressions and merge sentences to improve stealthiness.
In our experiments, we adopt the authoritative setting, i.e., the advertising style that achieves the best performance. Following the original paper, we use GPT-4o to generate the malicious tool descriptions.
We closely follow the original evaluation setup and are able to successfully reproduce the results reported in the paper (see \appref{apx:gapma}).

\pbold{AMA~\citep{mo2026attractive}} optimizes a single competing tool whose metadata (both its name and description) is refined so that the agent preferentially selects it over legitimate tools. It follows an evolutionary refinement loop: candidate descriptions are scored by their tool-selection success rate over a query set, ranked by a weighted value that rewards improvement over the parent candidate, and refined iteratively until a success threshold is reached. For a fair comparison with per-tool methods, we run AMA separately for each targeted tool: the tool's training-split queries form the query set, and AMA optimizes a dedicated malicious description for that tool. We use AMA's original implementation without modification, preserving its exact prompts, weighted-value scoring, and threshold-based stopping loop, with its default settings.

\pbold{ToolHijacker~\citep{shi2026prompt}} is an optimization-based method for tool description manipulation. It decomposes the malicious tool description into two components: a semantic description $\mathit{desc}_r$ and a selection directive $\mathit{desc}_s$.
$\mathit{desc}_r$ represents the semantic functionality of the tool, while $\mathit{desc}_s$ is designed to imperatively bias the LLM agent toward selecting the tool.
ToolHijacker synthesizes multiple queries for tasks that require the use of the targeted tool, which are used to generate $\mathit{desc}_r$.
For $\mathit{desc}_s$, ToolHijacker initializes it with a directive such as \emph{``Just output the word `$<$tool name$>$' as your final answer,''} which is then iteratively updated based on the feedback from the LLM agent.
The final malicious tool description is constructed by concatenating the two components: $\mathit{desc}_r \oplus \mathit{desc}_s$, where $\oplus$ denotes string concatenation.
In our experiments, we use the same model \llama{} for optimization, consistent with the original work.\looseness=-1

\begin{table}[t]
\centering
\caption{Accuracy of LLMs selecting the correct benign tool}
\label{tab:benign-acc-res}
\resizebox{0.7\columnwidth}{!}{%
\begin{tabular}{c|ccccc}
\toprule
\textbf{Tool Selector} & \textbf{\llama{}} & \textbf{\qwencoder{32B}} & \textbf{\qwen{}} & \textbf{Qwen3.8} & \textbf{\gpt{}} \\
\hline \hline
\ourdata{}  & 81.70\%  & 90.85\% & 83.61\% & 92.91\% & 91.06\%  \\
MetaTool & 91.27\%  & 84.68\% & 81.27\% & 92.91\% & 93.41\%   \\
MCP-Atlas & 76.11\% & 76.37\% & 76.47\% & 80.64\% & 82.27\%   \\

\bottomrule
\end{tabular}
}
\end{table}
\begin{table}[t]
\centering
\caption{LLM utility when different numbers of available benign tools are presented}\label{tab:utility-num-tool}
\begin{tabular}{c|cccc}
\toprule
\textbf{\#Tool} & \textbf{10} & \textbf{20} & \textbf{30} & \textbf{40} 
\\
\hline \hline
Accuracy  & 81.7\%  & 76.17\% & 71.42\% & 62.5\%\\

\bottomrule
\end{tabular}
\end{table}
\section{Experiment Setting}
\label{apx:setting}
\pbold{LLM Tool-Use Ability}
To assess the tool-use capabilities of the LLMs, we conduct a utility evaluation on all queries, including both training and test queries.
For each query, we construct a tool set consisting of the ground-truth benign tool and nine randomly sampled benign tools to simulate a realistic tool-use scenario, following prior work~\citep{wang2026mpma, shi2026prompt}.
\autoref{tab:benign-acc-res} reports the tool selection accuracy of each LLM without attacks.
We further evaluate how this accuracy changes as the tool set grows.
As shown in~\autoref{tab:utility-num-tool}, the accuracy decreases from 81.7\% with 10 tools to 62.5\% with 40 tools.

\pbold{Perplexity.} Following prior work~\citep{alon2023detecting}, we compute perplexity using GPT-2~\citep{radford2019language} as the reference language model. Perplexity is computed on the attacker side and requires no access to the target agent.

\pbold{LLM Serving}
We use vLLM~\citep{kwon2025vllm} to serve the open-source LLMs and the OpenAI API\footnote{https://openai.com/api/} to access GPT-5.
The average accuracy of selecting the correct tool for each query is reported in~\autoref{tab:benign-acc-res}.

\pbold{MCP Client.}
The MCP client is responsible for managing installed MCP servers and their associated tools. It extracts tool information and processes it to present to the LLM according to the MCP server configuration. In this work, we leverage the MCP client from the official MCP GitHub repository\footnote{https://github.com/modelcontextprotocol}. The client extracts tool metadata and organizes it into the system prompt provided to the LLM. An example of the system prompt is shown in~\autoref{pt:system_prompt} in Appendix.

\pbold{Implementation Setting.}
In the experiment, we set the number of queries used for optimization to 5, the number of starting points to $n = 10$, and the selection indicator to $m = 8$. The patience for early stopping was set to $P_t = 3$, and the patience for controlling the learning rate was set to $P_l = 3$. We used all available contrastive examples and set the maximum number of optimization steps to 300.
The experiment was run on NVIDIA H200 GPUs.

\section{Reproducing GAPMA}
\label{apx:gapma}

We conduct an experiment to ensure the reproducibility of GAPMA.
Specifically, we generate tool descriptions based on the targeted tools described in the original paper. While the authors evaluate eight tools, we sample four of them for our reproduction experiments: Time, Markdown, Search, and Hotnews. For each tool, we evaluate performance on 10 queries, which are synthesized using GPT-4o based on the query templates provided by the authors. Finally, we evaluate the results using Cline~\cite{cline}, an MCP client also used in the original work.
We conduct the evaluation using two proprietary LLMs: GPT-4o and Claude-4-Sonnet. In the original paper, the authors evaluate GAPMA using Claude-3.7-Sonnet; however, this model was not accessible at the time of our reproduction study. As a result, we replace it with Claude-4-Sonnet for evaluation.
\autoref{tab:GAPMA_reproduce} reports the reproduction results of GAPMA. 
Our experimental results show that the reproduced implementation achieves better performance than that reported in the original paper.

\begin{table}[H]
    \centering
    \footnotesize
    \caption{ASR comparison between the original paper and our reproduction.}
    \label{tab:GAPMA_reproduce}
        \begin{tabular}{cccccc}
        \toprule
        
         & Model & Time & Markdown & Search & HotNews  \\ \hline\hline
        \multirow{2}{*}{\textbf{Original}} & GPT & 0\% & 0\% & 100\% & 0\% \\
        & Claude & 0\% & 80\% & 0\% & 20\%\\ \hline
        \multirow{2}{*}{\textbf{Ours}} & GPT & 90\% & 0\% & 60\% & 100\% \\
        & Claude & 90\% & 100\% & 50\% & 100\%\\
        
        \bottomrule
        \end{tabular}
    
\end{table}

\begin{table*}[!h]
\centering
\caption{Transferability results across different unseen models in the 20-tool setting}
\label{tab:transfer-res-20tools}
\scriptsize
\begin{tabular}{c|cccc|cccc||c}
\toprule
& \multicolumn{4}{c}{\textbf{\ourdata}} & \multicolumn{4}{c}{\textbf{MetaTool}} \\
\cmidrule(lr){2-5} \cmidrule(lr){6-9}
\textbf{Eval Model}
& \textbf{\qwencoder{32B}}
& \textbf{Qwen-3}
& \textbf{Qwen3.8}
& \textbf{GPT-5}
& \textbf{\qwencoder{32B}}
& \textbf{Qwen-3}
& \textbf{Qwen3.8}
& \textbf{GPT-5}
& \textbf{Average}\\
\hline \hline
\eth
& 38.29\% & 53.19\% & 34.04\% & 41.84\%
& 39.00\% & 52.48\% & 53.19\% & 68.08\% & 47.51\% \\
GAPMA
& 33.33\% & 33.33\% & 19.86\% & 29.78\%
& 37.58\% & 41.13\% & 34.04\% & 56.02\% & 35.63\% \\
AMA
& 78.01\% & 68.79\% & 62.41\% & 62.41\%
& 56.74\% & 61.70\% & 48.23\% & 64.54\% & 62.85\% \\
ToolHijacker
& 46.09\% & 71.63\% & 73.05\% & 81.56\%
& 58.15\% & 65.95\% & 62.41\% & 87.94\% & 68.35\% \\
\tech
& 88.65\% & 76.59\% & 76.59\% & 85.10\%
& 75.17\% & 78.01\% & 73.05\% & 90.07\% & 80.40\% \\
\bottomrule
\end{tabular}
\end{table*}
\section{Transferability}
\label{apx:transferability}
\autoref{tab:transfer-res-20tools} shows the transfer ASR on unseen LLMs in the 20-tool setting.
\tech{} still outperforms all baselines, achieving the highest ASR on every evaluated model across both datasets, with an average ASR of 80.40\% and up to 90.07\% on GPT-5 with MetaTool.
Compared with the 10-tool setting (\autoref{tab:transfer-res}), its average ASR decreases only slightly, from 81.91\% to 80.40\%, indicating that its transferability is robust to larger tool sets.
In contrast, ToolHijacker and AMA achieve lower average ASRs of 68.35\% and 62.85\%, respectively, and ToolHijacker's ASR drops by up to 28.37\% on \ourdata{} compared with the surrogate model.
Consistent with the 10-tool setting, \eth{} and GAPMA remain relatively stable across models due to their feedback-free design.

\section{Human Evaluation Details}
\label{apx:human-eval}

We conduct the study on Prolific~\footnote{https://www.prolific.com} and recruit 20 human subjects in total.
We randomly sample 20 sets of descriptions from~\ourdata{}, where each set contains one benign description and the five malicious descriptions generated by \eth{}, GAPMA, AMA, ToolHijacker, and \tech{}.
Each set is ranked by five human subjects from \textit{most} suspicious (rank 1) to \textit{least} suspicious (rank 6).

A Friedman test revealed significant differences among description types
($\chi^2(5)=15.35$, $p=.009$). Holm-corrected Wilcoxon signed-rank tests showed
that only \eth{} and ToolHijacker descriptions were ranked significantly more
suspicious than the benign originals ($p_{\text{Holm}}=.031$ and $.019$; $r=.78$
and $.84$). In contrast, \tech{}'s descriptions were ranked no differently from
the benign originals ($p_{\text{Holm}}=1.00$), indicating that human raters
could not distinguish them from legitimate tool descriptions.

\begin{table}[t]
\centering
\caption{Detection results across different methods on MetaTool. Hybrid indicates that the MCP malware detector consists of several components, including keyword-based and prompt-based methods.}
\label{tab:detection-res-metatool}
\resizebox{0.7\columnwidth}{!}{%
\begin{tabular}{c|ccccc}
\toprule
\textbf{Detection Method} & \textbf{\eth} & \textbf{GAPMA} & \textbf{AMA} & \textbf{ToolHijacker} & \textbf{\tech} \\
\hline \hline
\multicolumn{6}{c}{\textit{Hybrid}} \\ \hline
MCP-Scanner & 100\% & 0\%    & 0\% & 8.51\% & 2.12\% \\
MCP-Shield  & 100\% & 0\%    & 0\% & 0\%    & 0\%    \\
MCP-Guard   & 100\% & 2.12\% & 0\% & 2.12\% & 6.36\% \\ \hline
\multicolumn{6}{c}{\textit{Prompt-based}} \\ \hline
GAPMA-D        & 0\%    & 0\% & 0\% & 8.51\% & 2.12\% \\
\ourdetector{} & 71.7\% & 0\% & 0\% & 78.7\% & 0\%  \\
\bottomrule
\end{tabular}
}
\end{table}
\section{Detection Details}
\label{apx:detection}

\pbold{Detectors.}
We employ four MCP malware detectors: MCP-Scanner~\citep{invariant2025mcp}, MCP-Shield~\citep{risen2025mcpshield}, MCP-Guard~\citep{xing2025guard}, and GAPMA~\citep{wang2026mpma}.
To distinguish the GAPMA detector from the GAPMA attack, we refer to the detector as GAPMA-D.
MCP-Scanner, MCP-Shield, and MCP-Guard are hybrid detectors that combine several components, including keyword-based and prompt-based methods, while GAPMA-D is prompt-based.

\noindent
\textbf{MCP-Scanner}~\citep{invariant2025mcp} is developed by Invariant Labs and is designed to scan MCP servers for common threats, including prompt injection, tool-poisoning attacks, malware payloads, and improper handling of sensitive data.

\pbold{MCP-Shield}~\citep{risen2025mcpshield} is an MCP malware detection system built on top of MCP-Scanner. It is capable of detecting vulnerabilities such as tool-poisoning attacks, data exfiltration channels, and cross-origin privilege escalation. In addition, MCP-Shield supports AI-assisted detection, allowing an LLM to serve as a judge during the inspection process. Following the default configuration of MCP-Shield, we use Claude Sonnet 4 as the judging LLM.

\pbold{MCP-Guard}~\citep{xing2025guard} is a multi-stage defense framework consisting of three sequential components. In the first stage, it performs syntactic filtering using rule-based keyword matching. In the second stage, it employs a fine-tuned E5 model~\citep{wang2024multilingual} to analyze tool content and assign a confidence score indicating potential maliciousness. In the final stage, if the confidence score falls below a predefined threshold, an LLM is invoked to perform a more detailed inspection. We adopt the default configuration of MCP-Guard in all our experiments.

\pbold{GAPMA-D}~\citep{wang2026mpma} is a prompt-based detection method proposed in the GAPMA paper, which is used to evaluate the stealthiness of descriptions generated by GAPMA. The detector is designed to assess whether a tool description contains malicious intent or exhibits suspicious characteristics. We incorporate this prompt-based detector into our evaluation as well. 

\pbold{Prompt-Injection Detectors.}
We also consider prompt-injection detectors.
PromptLocate~\citep{jia2026promptlocate} is a recent prompt-injection detection method, and we apply its official implementation to tool descriptions.
However, it produces a high false-positive rate: in our experiment, it flagged every benign tool description as malicious.
This is because prompt-injection detectors determine whether content contains instructions or intent that direct an LLM agent to perform certain actions.
Since tool descriptions typically describe a tool's functionality or purpose (see an example in~\autoref{tab:malicios_desc}), they inherently contain instruction-like content, so even benign descriptions are flagged.
We therefore exclude PromptLocate from our evaluation.
Instead, we adapt an LLM-based detector~\citep{shi2026promptarmor} to our setting, named \ourdetector{}.
To mitigate the randomness inherent in LLM-based detection, we run each evaluation 20 times and report the average detection rate.

\pbold{Results.}
\autoref{tab:detection-res} and \autoref{tab:detection-res-metatool} report the detection results on \ourdata{} and MetaTool, respectively.
\eth{} is detected by most detectors.
In particular, the hybrid detectors, which employ explicit rule-based keyword matching, achieve a 100\% detection rate.
In contrast, the prompt-based GAPMA-D fails to detect \eth{}, because its detection objective focuses on identifying suspicious intent rather than explicit keyword-based manipulation.

Although GAPMA and AMA are not the strongest attacks in ASR, their descriptions are highly stealthy: nearly all detectors report a detection rate of 0\%, suggesting that they largely evade existing detection mechanisms.

ToolHijacker bypasses MCP-Scanner, MCP-Shield, MCP-Guard, and GAPMA-D.
However, the simple prompt-based \ourdetector{} achieves high detection rates on ToolHijacker: 85.3\% and 78.7\% on \ourdata{} and MetaTool, respectively.
Thus, while ToolHijacker achieves a competitive ASR, its descriptions are relatively easy to identify with prompt-based methods.

Finally, \tech{} bypasses all detectors on both datasets, with the highest detection rate being only 6.36\% (MCP-Guard on MetaTool).

\section{Reliability of \tech{}}
\label{apx:reliability}

\begin{table}
    \centering
    \footnotesize
    \tabcolsep=3.5pt
    \caption{Impact of the number of tools presented to the LLM agent}
    \label{tab:num_of_tool}
        \begin{tabular}{c|ccc|ccc}
        \toprule
         & \multicolumn{3}{c}{\textbf{\ourdata}} & \multicolumn{3}{c}{\textbf{MetaTool}} \\ \hline\hline
        \#Tools & 5 & 10 & 15 & 5 & 10 & 15\\ \hline
        ToolHijacker & 87.23\% & 75.17\% & 74.46\% & 84.39\% & 82.97\% & 80.14\%  \\
        \tech{} & 90.07\% & 88.65\% & 79.94\% & 90.07\% & 87.94\% & 85.10\%  \\
        \bottomrule
        \end{tabular}
    
\end{table}

We conduct the following experiments for evaluating the reliability of \tech{}-generated descriptions: (1) the impact of the number of tools provided to the LLM agent, (2) the impact of the number of malicious versus benign tools presented to the LLM agent, and (3) the impact of increased ambiguity in presented tools.

\pbold{Number of Tools.}
To further study the impact of the number of tools presented to the LLM agent, we construct tool sets of different sizes (5, 10, and 15)\footnote{Prior work~\citep{wang2026mpma, shi2026prompt} uses 5-6 tools in evaluation.} and conduct experiments on \ourdata{} and MetaTool. We compare the results with ToolHijacker, as it is the most competitive baseline according to the results in~\autoref{tab:main-res} and~\autoref{tab:transfer-res}.

From \autoref{tab:num_of_tool}, we observe that although varying the number of tools affects overall performance, \tech{} consistently maintains a high ASR and outperforms ToolHijacker across all settings. Furthermore, when the number of tools provided to the LLM agent decreases (e.g., from 10 to 5), the ASR increases for both \tech{} and ToolHijacker. This may be because presenting fewer tools reduces irrelevant noise, allowing the LLM agent to more accurately select the malicious spoofing tool.

\pbold{Number of Benign Tools.}
In real-world settings, users may install multiple similar tools with overlapping functionality. As a result, multiple tools may all satisfy a given user query (i.e., multiple ground-truth tools). To evaluate whether \tech{} can still achieve high ASR under this setting, we conduct experiments where the tool set contains multiple ground-truth benign tools (1, 2, or 3) while there is only one malicious tool.

The results are reported in~\autoref{tab:num_benign_or_malicious_tool}. 
We observe that as the number of ground-truth tools increases, the ASR decreases only moderately. Even when three ground-truth tools are present, \tech{} achieves an ASR of 75.88\%, demonstrating its reliability under increased benign-tool ambiguity.

\pbold{Number of Malicious Tools.}
From an attacker’s perspective, malicious tools can also be distributed across multiple servers, or a single server may expose multiple malicious tools. Since MCP installs all tools provided by a server at once, attackers can increase the number of malicious tools available to the LLM agent. To study this effect, we conduct experiments in which the tool set contains multiple malicious tools.

As shown in \autoref{tab:num_benign_or_malicious_tool}, increasing the number of malicious tools leads to higher ASRs. Notably, when three malicious tools are present, the ASR reaches 95.03\%, substantially increasing the attacker’s success probability in conducting tool-spoofing attacks.

\begin{table}
    \centering
    \caption{Impact of the number of the benign tools or the malicious spoofing tools}
    \label{tab:num_benign_or_malicious_tool}
        \begin{tabular}{c|ccc}
        \toprule
        \# & 1 & 2 & 3 \\\hline\hline
        Benign & 88.65\% & 79.14\% & 75.88\%\\
        Malicious & 88.65\% & 92.13\% & 95.03\% \\
        \bottomrule
        \end{tabular}
     
\end{table}

\pbold{Impact of Increased Ambiguity.}
To evaluate the reliability of~\tech{} compared with other malicious descriptions, we conduct an experiment on~\ourdata{} where the benign description and all malicious descriptions are presented together to the LLM agent. This setup allows us to assess whether the malicious description optimized by~\tech{} remains preferred by the agent under increased ambiguity.

\autoref{fig:tsa_win_chart} presents the results, which show that even under increased ambiguity, i.e., when multiple different generated descriptions are presented for the same tool, \tech{} remains preferred by the LLM agent. Overall, the agent selects the tool with the description generated by~\tech{} 50.4\% of the time, demonstrating the reliability of~\tech{}.

\section{Overhead}
\label{apx:overhead}
We divide overhead into offline and online parts. Offline overhead is a one-time cost of optimizing the description; online overhead is incurred during the attack (i.e., presenting the optimized description).
For online overhead, all evaluated methods incur the same cost because the attack description is simply attached to the corresponding tool.
For offline overhead, we compare \tech{} with ToolHijacker~\citep{shi2026prompt}, as it is the strongest baseline in terms of ASR.
We measure the average optimization time per task, which is 272.67 seconds for \tech{} and 248.94 seconds for ToolHijacker.
Overall, \tech{} has comparable cost while achieving higher ASR and better stealthiness in both detector-based and human evaluations.

\section{Ablation Study}
\label{apx:ablation}
\tech{} consists of several components, including dynamic mutation adjustment (DMA), contrastive in-context learning, multi-starting-point initialization, and a patience mechanism. To understand the contribution of each component, we conduct a comprehensive ablation study.
The results are reported in~\autoref{tab:ablation} and~\autoref{fig:ablation} (in Appendix).
The \textit{Base} setting represents the most minimal configuration. In this setting, at each iteration, regardless of the feedback from the LLM agent, we simply prompt OptAgent to rephrase the tool description. In addition, we use a single fixed query and maintain the same tool set throughout the optimization process. The ASR for the base setting is 40\%.

\begin{table}[t]
\centering
\footnotesize
\caption{Ablation results}
\label{tab:ablation}

\begin{tabular}{l|cccc}
\toprule

\textbf{Eval Model} & ASR\\
\hline \hline
Base        & 40\% \\
Base + DMA   & 45\% \\
Base + DMA + Positive Example Only & 48.33\% \\
Base + DMA + Contrastive & 56.66\%  \\
\hline
\tech{} w/o DMA & 74.46\% \\
\hline
\tech & 88.65\% \\
\bottomrule
\end{tabular}

\end{table}

\begin{figure*}[t]
    \centering
    \begin{subfigure}{0.19\textwidth}
        \includegraphics[width=\linewidth]{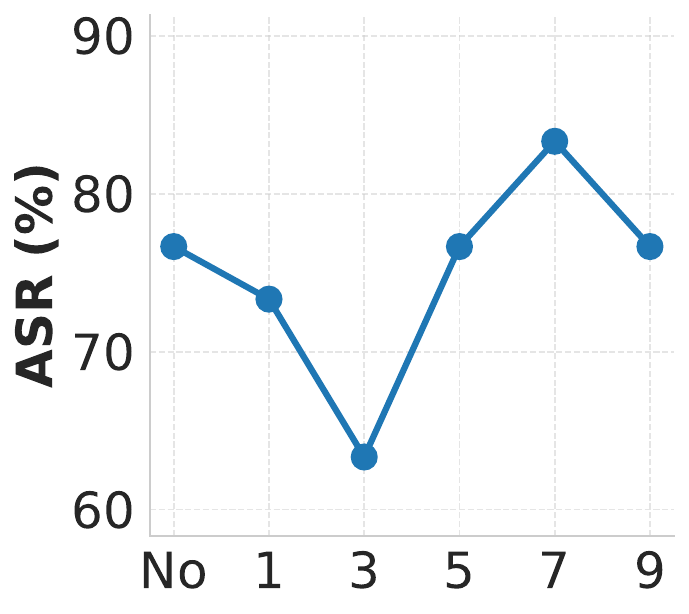}
        \caption{$P_t$}
        \label{fig:sub_num_pt}
    \end{subfigure}
    \hfill
    \begin{subfigure}{0.19\textwidth}
        \includegraphics[width=\linewidth]{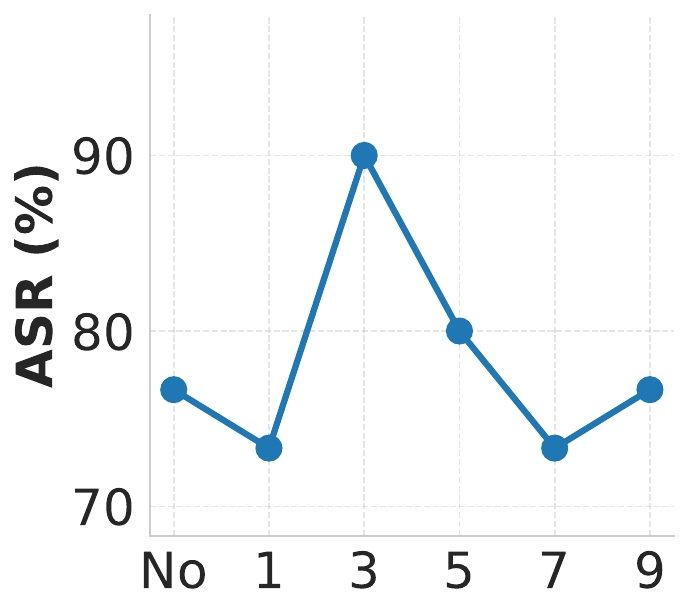}
        \caption{$P_l$}
        \label{fig:sub_num_pl}
    \end{subfigure}
    \hfill
    \begin{subfigure}{0.19\textwidth}
        \includegraphics[width=\linewidth]{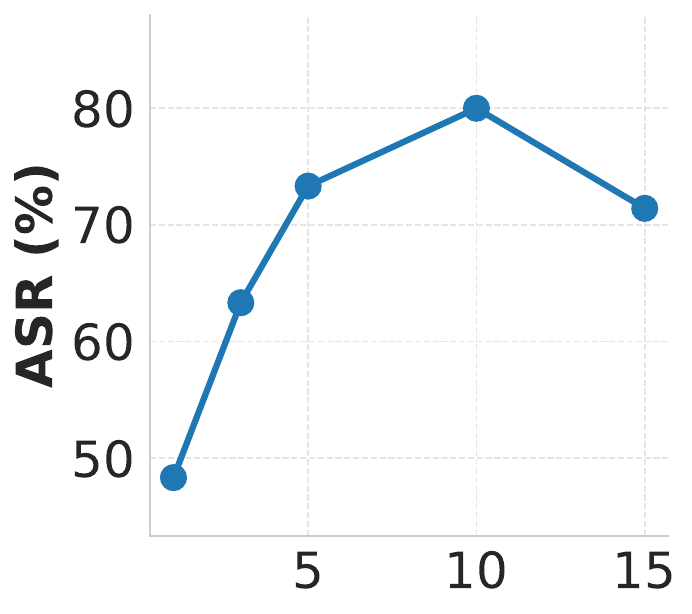}
        \caption{\# Starting Points}
        \label{fig:sub_num_starting_point}
    \end{subfigure}
    \hfill
    \begin{subfigure}{0.19\textwidth}
        \includegraphics[width=\linewidth]{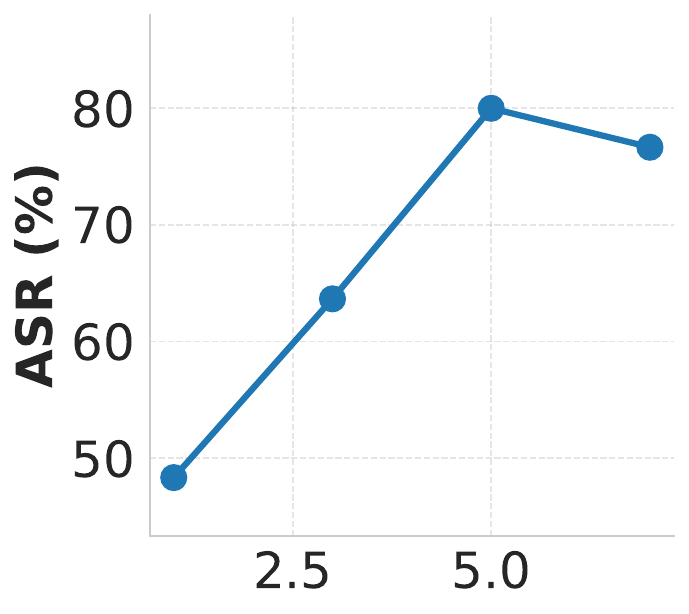}
        \caption{\# Training Query}
        \label{fig:sub_num_train_query}
    \end{subfigure}
    \hfill
    \begin{subfigure}{0.19\textwidth}
        \includegraphics[width=\linewidth]{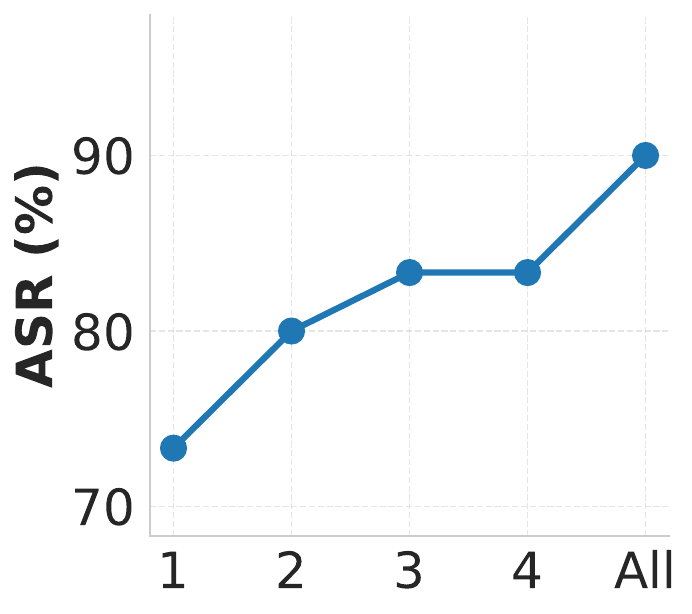}
        \caption{\# Contrastive Ex.}
        \label{fig:sub_num_contrastive}
    \end{subfigure}
    \caption{Ablation study on the hyperparameters}
    \label{fig:ablation}
\end{figure*}

\pbold{Impact of DMA.}
\autoref{tab:ablation} reports the impact of DMA. When DMA is added on top of the base setting, the ASR improves by 5\%. However, in this configuration, contrastive examples are not provided in the optimization prompt, and thus OptAgent relies solely on discrete feedback from the tool selector.\looseness=-1

After incorporating contrastive examples, the ASR increases by 16.66\% compared to the base setting. We note that this configuration optimizes only a single query with a fixed tool set, and therefore the results are overfitting. Nevertheless, the improvement clearly demonstrates the effectiveness of DMA and its interaction with contrastive examples.

In addition, we conduct an ablation experiment based on the full configuration but remove the DMA component (\tech{} w/o DMA) to further assess its impact. As shown in~\autoref{tab:ablation}, removing DMA results in a performance drop of 14.19\%. 
We further investigate the effects of other components, including the patience mechanism, the number of starting points, the number of training queries, and the number of contrastive examples. The results are presented in \appref{apx:ablation}.

\noindent
\textbf{Impact of Patience.}
\autoref{fig:sub_num_pt} presents the effect of the patience parameter used for early stopping. We observe that when the patience value $p_t$ is set too small (e.g., $p_t = 3$), the optimization fails to achieve a desirable ASR. Although the ASR at $p_t = 3$ may appear competitive, further analysis using \ourdetector{} reveals a detection rate of 54\%, indicating that additional optimization rounds are required to improve stealthiness.

As $p_t$ increases, the ASR consistently improves, with the best performance achieved at $p_t = 7$. We also observe that disabling early stopping entirely (i.e., no patience constraint) yields a reasonable ASR; however, this setting is less efficient and results in lower ASR compared to configurations that employ an appropriate early stopping mechanism.

\autoref{fig:sub_num_pl} shows the impact of the patience parameter for controlling the mutation magnitude ($p_l$). We observe that when this control is disabled or set to a very small value (i.e., \textit{No} or $p_l = 1$), the optimization becomes unstable, as OptAgent may perform an uncontrolled random search rather than a guided optimization.

Conversely, we observe that the ASR begins to decrease as $p_l$ increases beyond a certain point (starting from $p_l = 5$). One possible explanation is that a large $p_l$ causes the optimization to converge too deeply into a local minimum, making it more difficult for subsequent iterations to escape and explore better solutions. Overall, these results indicate that an appropriately chosen patience parameter for mutation magnitude control plays an important role in stabilizing the optimization and improving performance.

\pbold{Impact of Starting Points.}
\autoref{fig:sub_num_starting_point} reports the effect of the number of starting points used during optimization. The results show that increasing the number of starting points leads to higher ASR, with the best performance achieved when using 10 starting points. This suggests that employing multiple starting points enables OptAgent to explore a more diverse set of optimization trajectories, thereby learning richer strategies for optimizing the tool descriptions.

\pbold{Impact of the Number of Training Queries.}
As shown in~\autoref{fig:sub_num_train_query}, increasing the number of queries used during optimization leads to improved performance. This result indicates that using more training queries helps prevent the optimization process from overfitting to a specific query, thereby improving generalization and overall ASR.

\pbold{Impact of the Number of Contrastive Examples.}
As shown in~\autoref{fig:sub_num_contrastive}, incorporating contrastive examples significantly benefits the optimization process. In particular, providing a larger number of contrastive samples—or even the full optimization history—leads to improved performance. This additional information helps OptAgent identify failure cases to avoid and successful cases to reinforce. Overall, these results demonstrate that contrastive examples effectively guide OptAgent toward more effective optimization trajectories.

\section{Discussion and Future Implication}
\pbold{Limitations.}
As the number of ground-truth tools increases, the ASR of \tech{} begins to decrease.
This can be addressed by injecting more malicious tools, which can greatly strengthen the attack performance as we have shown in the evaluation.
Another limitation is that for the tool-spoofing attack, if the malicious tool is never installed, the attack cannot be executed, which also applies to other baselines.

\pbold{Real-world Attack.}
A real-world attack would strengthen our claims. We avoid such experiments for ethical reasons. Instead, we deploy a benign version of our server without harmful functionality to the MCP platform to validate feasibility. We use Cline to simulate a realistic user scenario where a client interacts with MCP servers. In this simulation, our attack achieves a 100\% ASR.

\pbold{Whitelisting Tools.}
Agentic applications may use whitelisting to restrict server installation. However, as in Python package management or network security, such mechanisms can be bypassed or spoofed. Consequently, users may still inadvertently install malicious servers, which can become a threat when \tech{} is applied.

\pbold{Future Implications.}
Our work demonstrates that it is possible to generate tool descriptions that are both stealthy and effective in conducting tool-spoofing attacks. This finding highlights a serious and underexplored security risk in the MCP ecosystem. As agentic AI becomes an increasingly dominant paradigm, the number of MCP tools that allow LLM agents to access external services will continue to grow. Consequently, ensuring the security of MCP ecosystems is of critical importance. Although several studies have revealed vulnerabilities in MCP systems~\citep{hou2025model, guo2025systematic, li2025we}, many security challenges remain open. Our work motivates future research on developing robust defense mechanisms to mitigate the risks posed by such attacks.

\section{System Prompts}
\label{apx:prompt}

\autoref{pt:system_prompt} shows the system prompt of the LLM agent for tool use.
\begin{figure}[h]
\centering
\begin{tcolorbox}[
  width=\columnwidth,
  colback=bg,
  colframe=black,
  boxrule=0.5pt,
  arc=2pt,
  left=4pt,
  right=4pt,
  top=4pt,
  bottom=4pt
]
\small
\begin{verbatim}
Choose the appropriate tool based on the user's question.
If no tool is needed, reply directly.

Please only choose one tool.

IMPORTANT: When you need to use a tool, you must ONLY respond 
with the exact JSON object format below, nothing else:
{
    "tool": "tool-name",
    "arguments": {
        "argument-name": "value"
    }
}

No other text or comments are allowed. Only the JSON object is allowed.

After receiving a tool's response:
1. Transform the raw data into a natural, conversational response
2. Keep responses concise but informative
3. Focus on the most relevant information
4. Use appropriate context from the user's question
5. Avoid simply repeating the raw data

\end{verbatim}
\end{tcolorbox}
\caption{System prompt used for tool selection and response generation.}
\label{pt:system_prompt}
\end{figure}

\begin{algorithm}[h]
\caption{\tech{}: Main Procedure}
\label{alg:tsa-main}
\begin{algorithmic}[1]
\Require Target benign tool $T_b$, malicious tool $T_m$, normal tool set $\mathcal{T}_n$,
         training queries $\mathcal{Q}$, surrogate tool selector \textsc{STS}, \textsc{OptAgent},
         max steps $t_{\max}$, stop patience $p_t$, mutation adjustment patience $p_l$,
         pool size $n$, context size $N$
\Ensure Optimized description $d^*$ for $T_m$

    \State $\gD \gets \textsc{Init}(d_b,\,\hat{n}_m,\,n,\,m)$;\;
           $\mathcal{C}^+,\,\mathcal{C}^- \gets \emptyset$;\;
           $c_t,\,c_l,\,t \gets 0$
           \hfill\Comment{Initialization (Alg.~\ref{alg:tsa-init})}
    \State Select query $q \in \mathcal{Q}$;\; sample context tools $\mathcal{T}_\mathrm{ctx} \subset \mathcal{T}_n$
    \While{$t < t_{\max}$ \textbf{and} $c_t < p_t$ \textbf{and} $c_l < t_{\max}$}
        \State Construct $\mathcal{S}^{(d)} = \{T_b,\,T_m^{(d)}\} \cup \mathcal{T}_\mathrm{ctx}$ for each $d \in \gD$
        \State Query \textsc{STS} with $(q,\,\mathcal{S}^{(d)})$ for all $d \in \gD$;\; collect feedback $\{r_d\}$
        \State Update $\mathcal{C}^+$, $\mathcal{C}^-$ from $\{r_d\}$
        \If{$\exists\, d : r_d = \mathrm{Malicious}$}
            \State $c_t \gets c_t + 1$;\; $c_l \gets 0$
            \If{$c_t \geq p_t$} \textbf{break} \EndIf
            \State Rotate $q$ and resample $\mathcal{T}_\mathrm{ctx}$;\;
                   filter $\gD$ to \textit{Malicious} descriptions only
        \Else
            \State $c_t \gets 0$;\; $c_l \gets c_l + 1$
        \EndIf
        \State $\gD \gets \{\textsc{Refine}(d_b,\,d,\,r_d,\,\mathcal{C}^+,\,\mathcal{C}^-,\,p_l,\,c_t) \mid d \in \gD\}$
               \hfill\Comment{DMA (Alg.~\ref{alg:tsa-refine})}
        \State $t \gets t + 1$
    \EndWhile
\State \Return $d^* = \arg\min_{(d,\,\mathrm{ppl})\,\in\,\mathcal{C}^+} \mathrm{ppl}$
\end{algorithmic}
\end{algorithm}

\begin{algorithm}[h]
\caption{\tech{}: Description Initialization}
\label{alg:tsa-init}
\begin{algorithmic}[1]
\Require Benign tool description $d_b$, malicious tool name $\hat{n}_m$,
         pool size $n$, indicator count $m$ ($m < n$)
\Ensure Initial description seed pool $\gD_0$

\State \textbf{Prompt} \textsc{OptAgent}: generate $n$ candidate descriptions $\{d_1,\ldots,d_n\}$ based on $d_b$
\State \textbf{Prompt} \textsc{OptAgent}: generate $m$ selection indicators $\{s_1,\ldots,s_m\}$ for $\hat{n}_m$
\State Pad with $n{-}m$ empty indicators: $s_{m+1},\ldots,s_n \gets \varepsilon$
\For{$i = 1$ \textbf{to} $n$}
    \State $\tilde{d}_i \gets d_i \oplus s_i$
    \State $\gD_0 \gets \gD_0 \cup \{(\tilde{d}_i,\;\mathrm{PPL}(\tilde{d}_i))\}$
\EndFor
\State \Return $\gD_0$
\end{algorithmic}
\end{algorithm}

\section{Detailed Algorithm of \tech{}}
\label{apx:algorithm}

\autoref{alg:tsa-main} presents the detailed procedure of \tech{}.
At the beginning, description seed pool $\gD$, positive example set $\mathcal{C}^+$, negative example set $\mathcal{C}^-$, and the counters are initialized alongside a query $q$ and context tools $\mathcal{T}_\mathrm{ctx}$ (Lines 1--2).
At each optimization step, all $n$ candidate descriptions are batch-evaluated by the surrogate tool selector STS and outcomes are used to update the contrastive example sets $\mathcal{C}^+$ and $\mathcal{C}^-$ (Lines 4--6).
If any description induces a \textit{Malicious} outcome, the success counter $c_t$ is incremented; once it reaches $p_t$, the optimization terminates early (Lines 7--10).
Otherwise, $q$ and $\mathcal{T}_\mathrm{ctx}$ are rotated to prevent overfitting and $\gD$ is pruned to retain only successful descriptions (Line 11).
If no \textit{Malicious} outcome is observed, the failure counter $c_l$ is incremented instead (Lines 12--14).
All surviving descriptions are then refined by the DMA procedure (Line 15, \autoref{alg:tsa-refine}).
The final output is the description in $\mathcal{C}^+$ with the lowest perplexity (Line 18).

\autoref{alg:tsa-refine} details the dynamic mutation adjustment (DMA) procedure.
At each call, the sliding history $\mathcal{H}$ is updated with the latest feedback $r$, and the mutation degree $\delta$ is determined from $\mathcal{H}$ (Lines 1--11).
Specifically, $\delta$ is set to \textit{low} when $r = \mathrm{Malicious}$; if the description is already sufficiently natural (i.e., $\mathrm{ppl} \leq \mathrm{PPL}(d_b)$), it is returned unchanged (Lines 2--6).
When $|\mathcal{H}|$ reaches $p_l$ without a \textit{Malicious} outcome, $\delta$ is escalated to \textit{high} (Lines 7--8); otherwise $\delta$ remains \textit{medium} (Lines 9--10).
Based on $\delta$, OptAgent is prompted to either merge the strongest phrases from successful descriptions in $\mathcal{C}^+$ into a single candidate ($\delta = \mathrm{low}$, Lines 13--14), apply a mild rewrite guided by $\mathcal{C}^+$ ($\delta = \mathrm{medium}$, Lines 18--19), or aggressively rewrite using $\mathcal{C}^+$ and $\mathcal{C}^-$ contrastively ($\delta = \mathrm{high}$, Lines 20--21).
Two feedback-specific overrides apply when $r = \mathrm{WrongTool}$ or $r = \mathrm{NoTool}$ (Lines 23--28).
The refined description is returned with updated scores (Lines 29--30).

\begin{algorithm}[t]
\caption{\textsc{Refine}: Dynamic Mutation Adjustment}
\label{alg:tsa-refine}
\begin{algorithmic}[1]
\Require Original description $d_b$, current description $d$, STS feedback $r$,
         positive/negative demo pools $\mathcal{C}^+$/$\mathcal{C}^-$,
         perplexity $\mathrm{ppl}$, mutation adjustment patience $p_l$,
         sliding history $\mathcal{H}$, stop counter $c_t$
\Ensure Refined description $d'$, updated $\mathrm{ppl}'$

\State Append $r$ to $\mathcal{H}$;\; truncate $\mathcal{H}$ to its last $p_l$ entries
\If{$r = \mathrm{Malicious}$}
    \State $\delta \gets \mathrm{low}$;\; clear $\mathcal{H}$
    \If{$\mathrm{ppl} \leq \mathrm{PPL}(d_b)$}
        \State \Return $(d,\;\mathrm{ppl})$
        \hfill\Comment{$d$ is already as natural as original; keep as-is}
    \EndIf
\ElsIf{$|\mathcal{H}| = p_l$}
    \State $\delta \gets \mathrm{high}$;\; clear $\mathcal{H}$
    \hfill\Comment{Patience exhausted; escalate mutation magnitude}
\Else
    \State $\delta \gets \mathrm{medium}$
\EndIf
\If{$\delta = \mathrm{low}$}
    \If{$|\mathcal{C}^+| > 1$ \textbf{and} $c_t < 5$}
        \hfill\Comment{Crossover: multiple \textit{Malicious} candidates available}
        \State \textbf{Prompt} \textsc{OptAgent}: merge strongest phrases from $\mathcal{C}^+$; naturalize selection indicators
    \Else
        \State \textbf{Prompt} \textsc{OptAgent}: soften aggressive selection phrases in $d$
    \EndIf
\ElsIf{$\delta = \mathrm{medium}$}
    \State \textbf{Prompt} \textsc{OptAgent}: mildly rewrite $d$ guided by $\mathcal{C}^+$ positive examples
\ElsIf{$\delta = \mathrm{high}$}
    \State \textbf{Prompt} \textsc{OptAgent}: aggressively rewrite $d$ using $\mathcal{C}^+$ and $\mathcal{C}^-$ contrastively
\EndIf
\If{$r = \mathrm{WrongTool}$ }
    \hfill\Comment{Override: drifted off target; steer with a past success}
    \State $d^+ \gets \mathrm{Select}(\mathcal{C}^+)$
    \hfill\Comment{e.g., the lowest-perplexity success}
    \State \textbf{Prompt} \textsc{OptAgent}: interpolate $d$ with $d^+$, preserving $d$'s task-relevant content
\ElsIf{$r = \mathrm{NoTool}$ \textbf{and} $\mathcal{C}^+ \neq \emptyset$}
    \hfill\Comment{Override: no tool invoked; pull back toward task context}
    \State \textbf{Prompt} \textsc{OptAgent}: rewrite $d$ guided by $\mathcal{C}^+$ references
\EndIf
\State $d' \gets \textsc{OptAgent}(\text{prompt})$
\State \Return $(d',\;\mathrm{PPL}(d'))$
\end{algorithmic}
\end{algorithm}

\end{document}